\documentclass[sn-mathphys,Numbered]{sn-jnl}% Math and Physical Sciences Reference Style
\usepackage{graphicx}%
\usepackage{multirow}%
\usepackage{amsmath,amssymb,amsfonts}%
\usepackage{amsthm}%
\usepackage{mathrsfs}%
\usepackage[title]{appendix}%
\usepackage{xcolor}%
\usepackage{textcomp}%
\usepackage{manyfoot}%
\usepackage{booktabs}%
\usepackage{algorithm}%
\usepackage{algorithmicx}%
\usepackage{algpseudocode}%
\usepackage{listings}%
\usepackage{tabularx}
\theoremstyle{thmstyleone}%
\theoremstyle{thmstyletwo}%

\theoremstyle{thmstylethree}%

\begin{document}

\title[article title]{Symmetry classification of temporal reciprocity in
time-varying electromagnetic media}


\author[1]{\fnm{Seulong} \sur{Kim}}

\author*[2,3]{\fnm{Kihong} \sur{Kim}}\email{khkim@ajou.ac.kr}

\affil[1]{\orgdiv{Research Institute of Basic Sciences}, \orgname{Ajou University}, \city{Suwon}, \postcode{16499}, \country{Korea}}

\affil[2]{\orgdiv{Department of Physics}, \orgname{Ajou University}, \city{Suwon}, \postcode{16499}, \country{Korea}}

\affil[3]{\orgdiv{School of Physics}, \orgname{Korea Institute for Advanced Study}, \city{Seoul}, \postcode{02455}, \country{Korea}}


\abstract
{
Time-varying electromagnetic media exhibit rich nonstationary wave phenomena, but the symmetry governing reversal of arbitrary temporal modulation sequences has remained unclear. We show that, in lossless, spatially homogeneous media with identical initial and final states, the scattering matrices of ordered and reversed sequences are related by inverse--conjugation, independent of the number of stages. This yields a classification of temporal reciprocity in bi-isotropic media: isotropic and chiral media are channel-preserving, whereas Tellegen media are channel-exchanging despite Lorentz nonreciprocity. Deterministic time rewinding follows directly. Our results provide a framework for predicting and designing temporal scattering responses in photonic media.
}

\maketitle

\section{Introduction}

Time-varying and spatiotemporally modulated photonic media have emerged as a fertile frontier in wave physics, enabling frequency conversion, nonadiabatic mode coupling, and time-domain analogs of mirrors and crystals \cite{gal,asg,morge}. Recent work has extended these concepts to magnetoelectric and bianisotropic systems, uncovering polarization- and direction-dependent scattering at temporal interfaces \cite{mir}. Yet despite this rapid progress, a fundamental question remains unanswered: what symmetry principle, if any, governs the reversal of arbitrary temporal modulation sequences, and how does this principle depend on the electromagnetic constitution of the medium?

This question is distinct from the classical problem of Lorentz reciprocity. In stationary systems, Lorentz reciprocity constrains source--field interchange and is broken by magnetoelectric media such as Tellegen materials \cite{tell1,caloz,asad}. Temporal or spatiotemporal modulation can similarly break spatial reciprocity, enabling magnetless nonreciprocal transmission \cite{estep,wang}. But Lorentz reciprocity governs spatial scattering, not temporal sequencing. Whether and how a wave system responds to reversal of its modulation history is a logically independent question, one that has not been systematically addressed. Time-reversal and time-mirror protocols have been explored
in various wave systems \cite{tv10,dr1}, and deterministic time
rewinding has been demonstrated in time-varying media \cite{rewind}.
Yet a general algebraic principle governing these phenomena
across material classes has remained elusive.

In spatially homogeneous media, temporal modulation at fixed wavevector defines a $2\times 2$ scattering problem in the frequency domain, with successive temporal interfaces and propagation intervals encoded in a scattering matrix. This formulation makes the symmetry question precise: what exact algebraic relation connects the scattering matrices of an ordered modulation sequence and its time-reverse? For lossless, spatially homogeneous media with identical initial and final states, we prove that, within this class, the two scattering matrices are always related by inverse--conjugation, independently of the number of modulation stages and valid for both propagating and evanescent intermediate intervals.

Crucially, the realization of this symmetry depends on electromagnetic material class. We show that time-varying bi-isotropic media fall into two distinct temporal reciprocity classes. Isotropic and chiral media obey a channel-preserving inverse--conjugation relation: temporal transmission is invariant under sequence reversal, and the pseudo-unitary $\mathrm{SU}(1,1)$ structure of lossless temporal scattering is preserved within each circular-polarization channel. Tellegen media, despite violating Lorentz reciprocity in the spatial sense, obey a polarization-exchanging variant: sequence reversal must be accompanied by exchange of circular-polarization channels, equivalently by reversal of the Tellegen parameter in the interface coefficients. Temporal reciprocity is therefore not destroyed by magnetoelectric nonreciprocity but reorganized into a distinct algebraic structure.

This classification has a direct and general corollary. Deterministic time rewinding, the exact reconstruction of time-evolving wave states in reversed temporal order, follows as a structural consequence of inverse--conjugation symmetry, rather than requiring specially engineered protocols. The conditions for time rewinding are fully determined by the material class, and we derive them explicitly for isotropic, chiral, and Tellegen media, including both propagating and evanescent modulation intervals. Recent experimental realization of microwave Tellegen metamaterials with giant magnetoelectric responses \cite{gtel} demonstrates that the polarization-exchanging reciprocity class identified here should be directly observable in time-modulated magnetoelectric platforms. This establishes a general symmetry principle for predicting how time-modulated electromagnetic systems respond to reversal of modulation history, including magnetoelectric platforms where temporal reciprocity takes a polarization-exchanging form.

\section{Results}
\subsection{Inverse--conjugation symmetry in isotropic and chiral media}

We establish the inverse--conjugation symmetry governing temporal reciprocity, beginning with
isotropic media and then extending the classification to the bi-isotropic family.
Throughout, $\epsilon$ and $\mu$ are taken to be real but are unrestricted in sign, thereby encompassing positive- and negative-index propagation as well as evanescent intervals. These parameters should be understood as idealized effective values at a given operating point; in passive realizations, negative-parameter regimes are generally associated with dispersion, which is discussed below.

We first clarify the scope of the formulation. The media considered here are spatially homogeneous,
so the wave vector \(k\) is conserved and temporal scattering can be analyzed independently in each
fixed-\(k\) sector. The exact symmetry relations derived below are formulated for lossless temporal
scattering with identical initial and final media. For clarity, we use a piecewise-constant temporal
representation, in which the evolution is written as a product of temporal interface and propagation
matrices. This representation is a convenient discretization, not a fundamental restriction: since the
inverse--conjugation relation is independent of the number of temporal intervals, continuously
varying modulations are obtained in the continuum limit.

Consider an ordered temporal sequence
\(\mathrm A\to\mathrm B\to\mathrm C\to\mathrm A\)
and its reversed counterpart
\(\mathrm A\to\mathrm C\to\mathrm B\to\mathrm A\).
Their transmission and reflection amplitudes satisfy
\begin{align}
s=\tilde{s},\qquad r=-\tilde r^{*},
\label{eq:eq1q}
\end{align}
where \(s,r\) \((\tilde{s},\tilde r)\) correspond to the ordered (reversed) sequence. Transmission is
invariant under temporal reversal, while reflection differs only by a universal phase inversion.
Figure~1 illustrates this channel-preserving reciprocity.

This symmetry is most transparently formulated in a temporal scattering formalism at fixed wave
vector \(\mathbf{k}\),
\begin{align}
\mathbf D_{\mathrm{out}}=\mathcal S\mathbf D_{\mathrm{in}},\qquad
\mathcal S=
\begin{pmatrix}
s & r'\\
r & s'
\end{pmatrix}.
\end{align}
Here \(\mathbf D_{\mathrm{in}}\) and \(\mathbf D_{\mathrm{out}}\) denote the incoming and outgoing
amplitudes of the two eigenfrequency branches associated with the same conserved wave vector.
In a lossless, spatially homogeneous medium, conservation of wave-action flux imposes the
pseudo-unitarity condition
\(\mathcal S^\dagger \Sigma \mathcal S=\Sigma\),
with \(\Sigma=\mathrm{diag}(1,-1)\). After removal of an overall phase, \(\mathcal S\) belongs to
\(\mathrm{SU}(1,1)\) and can be written in the canonical form
\begin{align}
\mathcal S=
\begin{pmatrix}
s & r^{*}\\
r & s^{*}
\end{pmatrix},
\label{eq_su11}
\end{align}
with
\(|s|^{2}-|r|^{2}=1\).
This form is closed under multiplication and is therefore preserved under arbitrary concatenation
of temporal scattering processes.

For a piecewise-constant temporal representation, an interface from medium \(\mathrm A\) to medium
\(\mathrm B\) is described by
\begin{align}
\mathcal S_{\mathrm A\to\mathrm B}=
\begin{pmatrix}
s_{\mathrm A\to\mathrm B} & r_{\mathrm A\to\mathrm B}\\
r_{\mathrm A\to\mathrm B} & s_{\mathrm A\to\mathrm B}
\end{pmatrix},
\end{align}
where
\begin{align}
s_{\mathrm A\to\mathrm B}
=\frac12\left(1+\frac{\eta_{\mathrm A}}{\eta_{\mathrm B}}\right),\qquad
r_{\mathrm A\to\mathrm B}
=\frac12\left(1-\frac{\eta_{\mathrm A}}{\eta_{\mathrm B}}\right),
\end{align}
and
\(\eta_i=\sqrt{\mu_i}/\sqrt{\epsilon_i}\)
is the wave impedance of medium \(i\). Propagation through a stationary interval of duration
\(\tau\) is represented by
\begin{align}
\mathcal P=
\begin{pmatrix}
e^{-i\phi} & 0\\
0 & e^{i\phi}
\end{pmatrix},
\qquad
\phi=\omega\tau,
\end{align}
with \(\omega=ck/n\) and \(n=\sqrt{\epsilon}\sqrt{\mu}\). Evanescent intervals are incorporated by
analytic continuation of \(\omega\) and \(\phi\).

We now summarize the derivation of the inverse--conjugation relation for isotropic media. Let
\(\mathrm I\) denote the identical initial and final medium, and write the contribution of temporal
interval \(j\) as the interface--propagation--interface block
\begin{equation}
Q_j=\mathcal S_{j\to \mathrm I}\mathcal P_j\mathcal S_{\mathrm I\to j},
\end{equation}
where \(\mathcal S_{\mathrm I\to j}\) and \(\mathcal S_{j\to \mathrm I}\) are temporal interface matrices and
\(\mathcal P_j\) is the propagation matrix accumulated in interval \(j\). An ordered sequence of
\(N\) intervals and its reversed counterpart can then be written as
\begin{equation}
\mathcal S_{\rm ord}
=
Q_NQ_{N-1}\cdots Q_1,
\qquad
\mathcal S_{\rm rev}
=
Q_1Q_2\cdots Q_N .
\end{equation}
The key point is that each block satisfies
\begin{equation}
Q_j=(Q_j^*)^{-1}.
\end{equation}
For a propagating interval, this follows from the interface inversion property
\(\mathcal S_{\mathrm I\to j}=\mathcal S_{j\to \mathrm I}^{-1}\) and the propagation identity
\(\mathcal P_j^{-1}=\mathcal P_j^*\).
For an evanescent interval, the propagation matrix satisfies
\(\mathcal P_j^{-1}=M\mathcal P_j M\), where \(M\) exchanges the two
eigenfrequency branches. In this case, complex conjugation of the temporal
interface matrices is also accompanied by the same branch exchange. Since the
block \(Q_j=\mathcal S_{j\to \mathrm I}\mathcal P_j\mathcal S_{\mathrm I\to j}\)
contains two interfaces, the branch-exchange matrices appear in pairs and
reduce through \(M^2=I\), leaving the same block identity
\(Q_j=(Q_j^*)^{-1}\). Thus the same result holds for both propagating and
evanescent intervals.

Taking the product of the block identities gives
\begin{equation}
\mathcal S_{\rm ord}
=
Q_NQ_{N-1}\cdots Q_1
=
\left(Q_1^*Q_2^*\cdots Q_N^*\right)^{-1}
=
\left(\mathcal S_{\rm rev}^*\right)^{-1}.
\end{equation}
This proves the inverse--conjugation symmetry
\begin{equation}
\mathcal S_{\rm ord}
=
\left(\mathcal S_{\rm rev}^{*}\right)^{-1},
\label{eq:conjrevw}
\end{equation}
for arbitrary temporal sequences with identical initial and final media, independently of the
number of intervals and valid for both propagating and evanescent intermediate intervals.

The full algebraic derivation, including the separate treatment of propagating and evanescent
intervals and the extension to arbitrary temporal sequences, is given in Supplementary Note~1.
The continuum-limit case is illustrated in Fig.~S3 of Supplement~1, which numerically verifies
the predicted temporal-reciprocity relations for a continuously varying chiral modulation with
smooth \(\epsilon(t)\), \(\mu(t)\), and \(\gamma(t)\).

Equation~(\ref{eq:eq1q}) follows directly from Eq.~(\ref{eq:conjrevw}). A direct corollary is
deterministic time rewinding, the exact reconstruction of time-evolving wave states in reversed
temporal order:
\begin{equation}
\mathcal S_{\rm rev}^{*}\mathcal S_{\rm ord}=I,
\end{equation}
implying exact restoration of the initial state for arbitrary inputs. The corresponding time-rewinding
conditions for isotropic media are derived in Supplementary Note~2. Isotropic media therefore
realize a channel-preserving inverse--conjugation symmetry, defining the first temporal reciprocity
class.

Importantly, temporal reciprocity operates at the level of the scattering matrix rather than at the
level of individual material segments: the reversed modulation need not retrace the ordered sequence
parameter by parameter. Durations and constitutive parameters may differ substantially, provided
the inverse--conjugation relation is satisfied.

\begin{figure*}[t]
  \centering
  \includegraphics[width=0.85\textwidth]{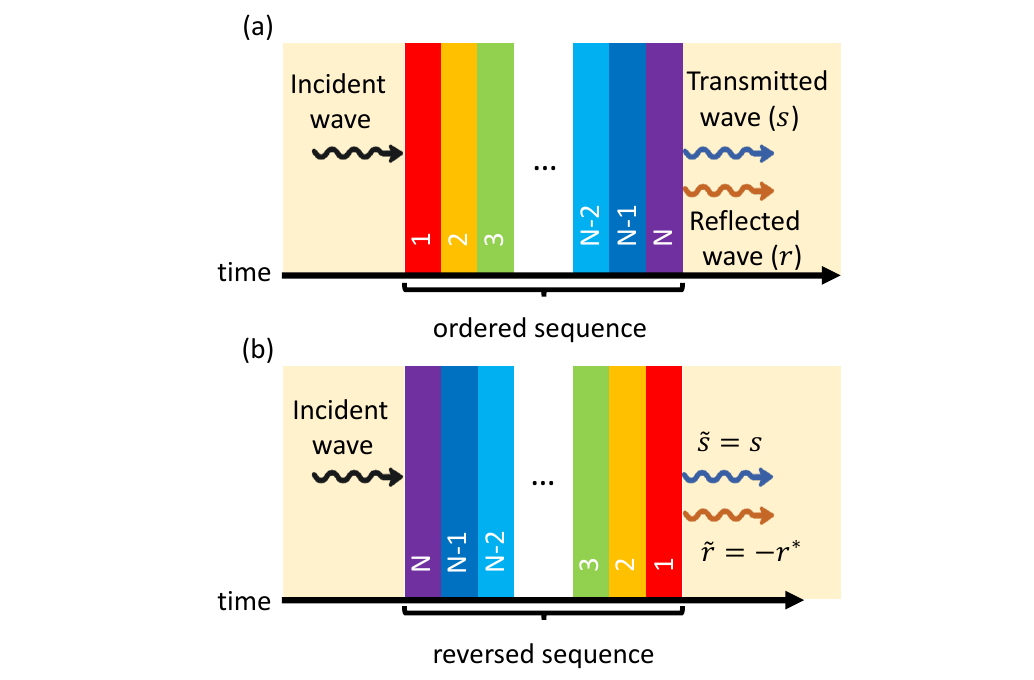}
  \caption{\textbf{Temporal reciprocity in time-modulated isotropic media.}
\textbf{a} An incident wave at fixed wavevector $k$ is scattered by an ordered temporal modulation sequence of $N$ slabs, producing transmitted and reflected components with amplitudes $s$ and $r$.
\textbf{b} Reversal of the temporal sequence, with identical initial and final media, yields a scattering matrix related to the original by inverse--conjugation. Transmission is invariant,
$\tilde s=s$, while reflection acquires a universal phase inversion, $\tilde r=-r^*$. Both relations hold for arbitrary $N$ and for propagating and evanescent intermediate intervals alike.
}
\end{figure*}

This classification extends naturally to bi-isotropic media, where the circular
basis diagonalizes the magnetoelectric coupling and exposes the relevant channel
structure. In a spatially homogeneous bi-isotropic medium, the constitutive
relations are
\begin{align}
\mathbf D=\epsilon\mathbf E+a\mathbf H,\qquad
\mathbf B=\mu\mathbf H+a^{*}\mathbf E,
\end{align}
where \(a=\chi+i\gamma\) combines the Tellegen parameter \(\chi\) and the
chirality parameter \(\gamma\). The limits \(\chi=0\) and \(\gamma=0\)
correspond to chiral and Tellegen media, respectively.

For plane waves with fixed wave vector \(\mathbf{k}=k\hat{\mathbf z}\),
Maxwell's equations decouple in the circular basis
\begin{align}
D_L=\frac{D_x+iD_y}{\sqrt{2}},\qquad
D_R=\frac{D_x-iD_y}{\sqrt{2}} .
\end{align}
The amplitudes \(D_L\) and \(D_R\) define two dynamically independent scalar
channels. The labels \(L\) and \(R\) denote these algebraically decoupled
sectors; by themselves they do not specify the direction of energy transport.
Within a stationary temporal interval, each channel supports two eigenfrequency
branches, \(\omega^{[L]}_{\pm}(k)\) and \(\omega^{[R]}_{\pm}(k)\), obtained from
the corresponding quadratic dispersion relations~\cite{tell2}. The circular
eigenmodes and dispersion relations are summarized in Supplementary Note~3. For
fixed \(k>0\), the sign of \(\omega\) determines the direction of energy
transport, so the two branches within each channel correspond to
counter-propagating waves. The physical circular polarization of a propagating
mode is therefore determined jointly by the channel index and the sign of
\(\omega\).

The dispersion relations yield
\begin{align}
n_{R,L}=n_0\pm\gamma,\qquad
\eta_{R,L}=\frac{n_0\pm i\chi}{\epsilon},
\label{eq:eta_clean}
\end{align}
where
\begin{align}
n_0=
\begin{cases}
\pm\sqrt{\epsilon\mu-\chi^2}, & \epsilon\mu>\chi^2,\\[4pt]
i\sqrt{\chi^2-\epsilon\mu}, & \epsilon\mu<\chi^2,
\end{cases}
\label{eq:n0_clean}
\end{align}
and the sign of \(n_0\) selects the refractive-index branch. Thus chirality
\(\gamma\) splits the circular refractive indices, whereas the Tellegen
parameter \(\chi\) makes the effective impedance polarization dependent. A
structural feature crucial for temporal scattering is that temporal boundary
conditions couple the two eigenfrequency branches within each channel,
\(\omega_+\leftrightarrow\omega_-\), while the \(L\) and \(R\) sectors remain
dynamically independent. Temporal scattering at fixed \(k\) therefore reduces,
in each circular channel, to a \(2\times2\) scattering problem connecting its two
frequency branches.

\textit{Chiral media.---}
Chiral media \((\gamma\neq0,\chi=0)\) belong to the same temporal-reciprocity
class as isotropic media. Although chirality splits the refractive indices, the
effective impedances remain polarization independent,
$\eta_R=\eta_L=n_0/\epsilon=
\sqrt{\mu}/\sqrt{\epsilon}$.
Thus the temporal interface coefficients are identical to those of isotropic
media. The structural reason is that the Tellegen parameter \(\chi\), rather
than the chirality parameter \(\gamma\), enters the temporal matching
conditions. Chirality therefore does not affect temporal interface scattering;
it enters only through the propagation phases accumulated during stationary
temporal intervals.

The propagation matrices obey the same inversion identities as in isotropic
media for propagating intervals, and their chiral counterpart for evanescent
intervals,
\begin{align}
\mathcal P^{-1}&=\mathcal P^{*}
\qquad \text{(propagating interval)},\nonumber\\
\mathcal P^{-1}&=M\mathcal P^{*}M
\qquad \text{(chiral evanescent interval)},
\end{align}
with the same branch-exchange matrix \(M\) as above. Consequently, chiral
media preserve the same \(\mathrm{SU}(1,1)\) scattering structure and obey the
channel-preserving inverse--conjugation relation
\[
\mathcal S_{\mathrm{ord}}
=
\left(\mathcal S_{\mathrm{rev}}^{*}\right)^{-1}
\]
within each circular channel. Isotropic and chiral media therefore form a single
temporal-reciprocity class: they differ in the propagation phases accumulated by
the two circular channels, but not in the underlying symmetry algebra. The
detailed derivation of temporal reciprocity and the time-rewinding conditions in
chiral media is given in Supplementary Notes~4 and~5.

\begin{table*}[t]
\centering
\caption{\textbf{Symmetry classification of temporal reciprocity in spatially homogeneous bi-isotropic media.}
Bi-isotropic media separate into two distinct temporal reciprocity classes determined by their magnetoelectric constitution. Isotropic and chiral media obey a channel-preserving inverse--conjugation symmetry, in which sequence reversal enforces ${\mathcal S}_{\rm ord}=({\mathcal S}_{\rm rev}^*)^{-1}$
within each circular-polarization channel independently. Tellegen media realize a channel-exchanging variant, ${\mathcal S}_{\rm ord}^R=[({\mathcal S}_{\rm rev}^L)^{*}]^{-1}$, in which sequence reversal must be accompanied by exchange of circular-polarization channels. The parameter ratios and sign conditions required for deterministic time rewinding are given for each class.}
\label{table1}
\renewcommand{\arraystretch}{1.5}
\begin{tabularx}{\textwidth}{ >{\raggedright\arraybackslash\hsize=0.9\hsize}X  >{\raggedright\arraybackslash\hsize=1\hsize}X  >{\raggedright\arraybackslash\hsize=1.05\hsize}X  >{\raggedright\arraybackslash\hsize=1.05\hsize}X }

\toprule
 & \textbf{Isotropic} & \textbf{Chiral} \newline ($\gamma\ne 0,~\chi=0$) & \textbf{Tellegen} \newline ($\chi\ne 0,~\gamma=0$) \\
\midrule \addlinespace[1pt]

Channel structure & Single channel & Two decoupled channels & Two decoupled channels \\ \addlinespace[1pt]

Interface polarization
& No & No \newline ($\eta_R=\eta_L$) & Yes \newline ($\eta_R\neq\eta_L$) \\ \addlinespace[1pt]

Reciprocity relation & $S_{\rm ord}=(S_{\rm rev}^*)^{-1}$ &
$S_{\rm ord}^{j} =[(S_{\rm rev}^{j})^{*}]^{-1}$\newline $(j =R,L) $ &
$S_{\rm ord}^{R}=[(S_{\rm rev}^{L})^{*}]^{-1}$ \\ \addlinespace[3pt]

Time-rewinding ratios &
$\dfrac{\epsilon'}{\epsilon} =\dfrac{\mu'}{\mu}$, \newline $\dfrac{\tau^\prime}{\tau} =\left\vert\dfrac{n^\prime}{n}\right\vert $  &
$\dfrac{\epsilon'}{\epsilon}=\dfrac{\mu'}{\mu} =\dfrac{\gamma'}{\gamma}$,  \newline $\dfrac{\tau^\prime}{\tau} =\left\vert\dfrac{n^\prime}{n}\right\vert $ &
$\dfrac{\epsilon'}{\epsilon}=\dfrac{\mu'}{\mu} =\dfrac{\chi'}{\chi}$, \newline $\dfrac{\tau^\prime}{\tau} =\left\vert\dfrac{n_0^\prime}{n_0}\right\vert $ \\ \addlinespace[3pt]

Sign condition &
$\text{sgn}(\epsilon'/\epsilon)$ $  =\text{sgn}(\mu'/\mu) =-1$  &
\text{sgn}($\epsilon'/\epsilon$) $=\text{sgn}(\mu'/\mu) $ $=\text{sgn}(\gamma'/\gamma)=-1$  &
\text{sgn}($\epsilon'/\epsilon$) $=\text{sgn}(\mu'/\mu)$ $=\text{sgn}(\chi'/\chi)=-1  $ \\ \addlinespace[1pt]
\bottomrule
\end{tabularx}
\end{table*}

\subsection{Polarization-exchanging temporal reciprocity in Tellegen media}

\textit{Tellegen media.---}
Tellegen media \((\chi\neq0,\gamma=0)\) constitute a distinct temporal-reciprocity
class because the distinction already appears at temporal interfaces. In this
case, the effective impedances
$\eta_{R,L}=(n_0\pm i\chi)/\epsilon$
depend on circular polarization, so the interface coefficients differ between
the \(D_R\) and \(D_L\) channels. For a temporal interface
\(\mathrm A\to\mathrm B\), they can be written as
\begin{align}
s_{\mathrm A\to\mathrm B}^{R/L}
&=\frac12\left(1+\rho_{\mathrm A\to\mathrm B}
\mp i\nu_{\mathrm A\to\mathrm B}\right),\nonumber\\
r_{\mathrm A\to\mathrm B}^{R/L}
&=\frac12\left(1-\rho_{\mathrm A\to\mathrm B}
\pm i\nu_{\mathrm A\to\mathrm B}\right),
\label{eq:tellegen_interface}
\end{align}
where
\begin{align}
\rho_{\mathrm A\to\mathrm B}
=
\frac{\epsilon_{\mathrm B} n_{0\mathrm A}}
{\epsilon_{\mathrm A} n_{0\mathrm B}},
\qquad
\nu_{\mathrm A\to\mathrm B}
=
\frac{\epsilon_{\mathrm A}\chi_{\mathrm B}
-\epsilon_{\mathrm B}\chi_{\mathrm A}}
{\epsilon_{\mathrm A}n_{0\mathrm B}} .
\label{eq:rho_nu}
\end{align}
Here \(n_{0j}\) is defined in Eq.~(\ref{eq:n0_clean}) for medium \(j\).
The parameter \(\rho_{\mathrm A\to\mathrm B}\) represents the impedance-like
mismatch associated with the effective refractive indices of the two media,
whereas \(\nu_{\mathrm A\to\mathrm B}\) measures the change of the Tellegen
response across the interface. These coefficients follow from the temporal
boundary conditions in the circular basis: \(D_{R,L}\) is continuous, together
with the corresponding derivative combination containing \(\chi\). Thus
\(\chi\), unlike \(\gamma\), enters temporal matching directly. A detailed
derivation of Eqs.~(\ref{eq:tellegen_interface}) and~(\ref{eq:rho_nu}) is given
in Supplementary Note~3.

As a result, the channel-preserving inverse--conjugation symmetry of the
isotropic--chiral class no longer holds within a single circular channel.
Temporal reciprocity nevertheless survives in a reorganized form. Complex
conjugation of the Tellegen interface coefficients exchanges the two circular
channels: the conjugate of the \(R\)-channel interface has the algebraic
structure of the corresponding \(L\)-channel interface, and vice versa.
Consequently, for arbitrary temporal modulation sequences with identical initial
and final media,
\begin{align}
\mathcal S_{\mathrm{ord}}^{R}
=
\left[
\left(
\mathcal S_{\mathrm{rev}}^{L}
\right)^{*}
\right]^{-1},
\label{eq:tellegen_reciprocity}
\end{align}
and similarly with \(R\) and \(L\) interchanged. Here \(\mathcal S^{R}\) and
\(\mathcal S^{L}\) denote the \(2\times2\) temporal scattering matrices in the
dynamically decoupled \(D_R\) and \(D_L\) channels. Thus temporal reversal must
be accompanied by circular-channel exchange, which is algebraically equivalent
to reversing the sign of the Tellegen parameter \(\chi\) in the interface
coefficients. This relation is independent of the number of temporal stages and
remains valid for both propagating and evanescent intermediate intervals. The
full derivation is given in Supplementary Note~6.

Within each circular channel, the scattering matrix retains the pseudo-unitary
\(\mathrm{SU}(1,1)\) structure of Eq.~(\ref{eq_su11}). Equation
~(\ref{eq:tellegen_reciprocity}) then implies
\begin{align}
s^{R}=\tilde s^{L},\qquad
r^{R}=-\left(\tilde r^{L}\right)^{*}.
\label{eq:cesym}
\end{align}
Transmission is preserved under temporal reversal after channel exchange, while
reflection acquires the same symmetry-enforced phase inversion as in the
channel-preserving class. Figure~2 confirms this polarization-exchanging
symmetry numerically: the usual single-channel relations fail, but the
channel-exchanged relations are fully restored. Additional numerical checks are
provided in Supplementary Note~8.

The physical interpretation is significant. Tellegen media violate Lorentz
reciprocity in the spatial sense, and one might expect this nonreciprocity to
compromise temporal reversibility as well. Equation
~(\ref{eq:tellegen_reciprocity}) shows that this expectation is incorrect:
magnetoelectric nonreciprocity does not break temporal reciprocity, but
reorganizes its algebraic structure from a channel-preserving to a
channel-exchanging form. The two symmetry classes are therefore distinguished not
by whether temporal reciprocity holds, but by how it is realized. Deterministic
time rewinding remains achievable within the channel-exchanged framework, with
the explicit rewinding conditions summarized in Table~1 and derived in
Supplementary Note~7.

The complete symmetry classification of temporal reciprocity in isotropic,
chiral, and Tellegen media is summarized in Table~1. The two classes are
distinguished by whether temporal interface scattering is polarization
dependent, which determines whether the inverse--conjugation relation is channel
preserving or channel exchanging and fixes the material-class-dependent
conditions for deterministic time rewinding.

\begin{figure}
  \centering
  \includegraphics[width=9cm]{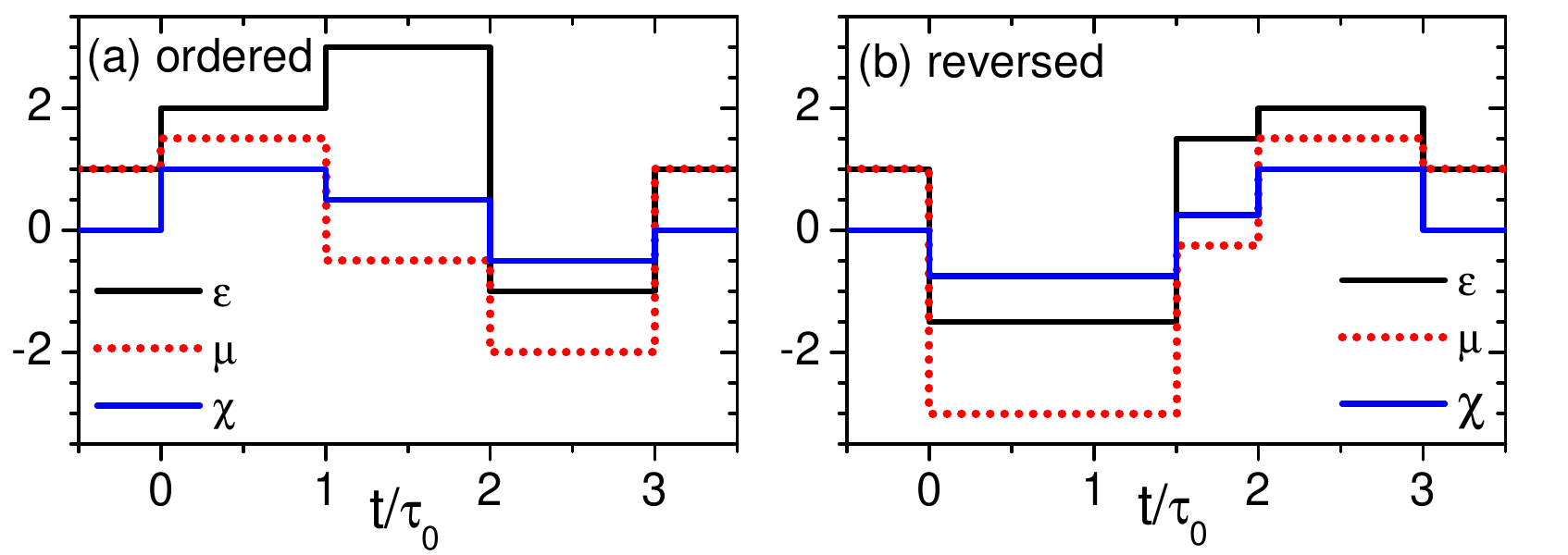}
  \includegraphics[width=9cm]{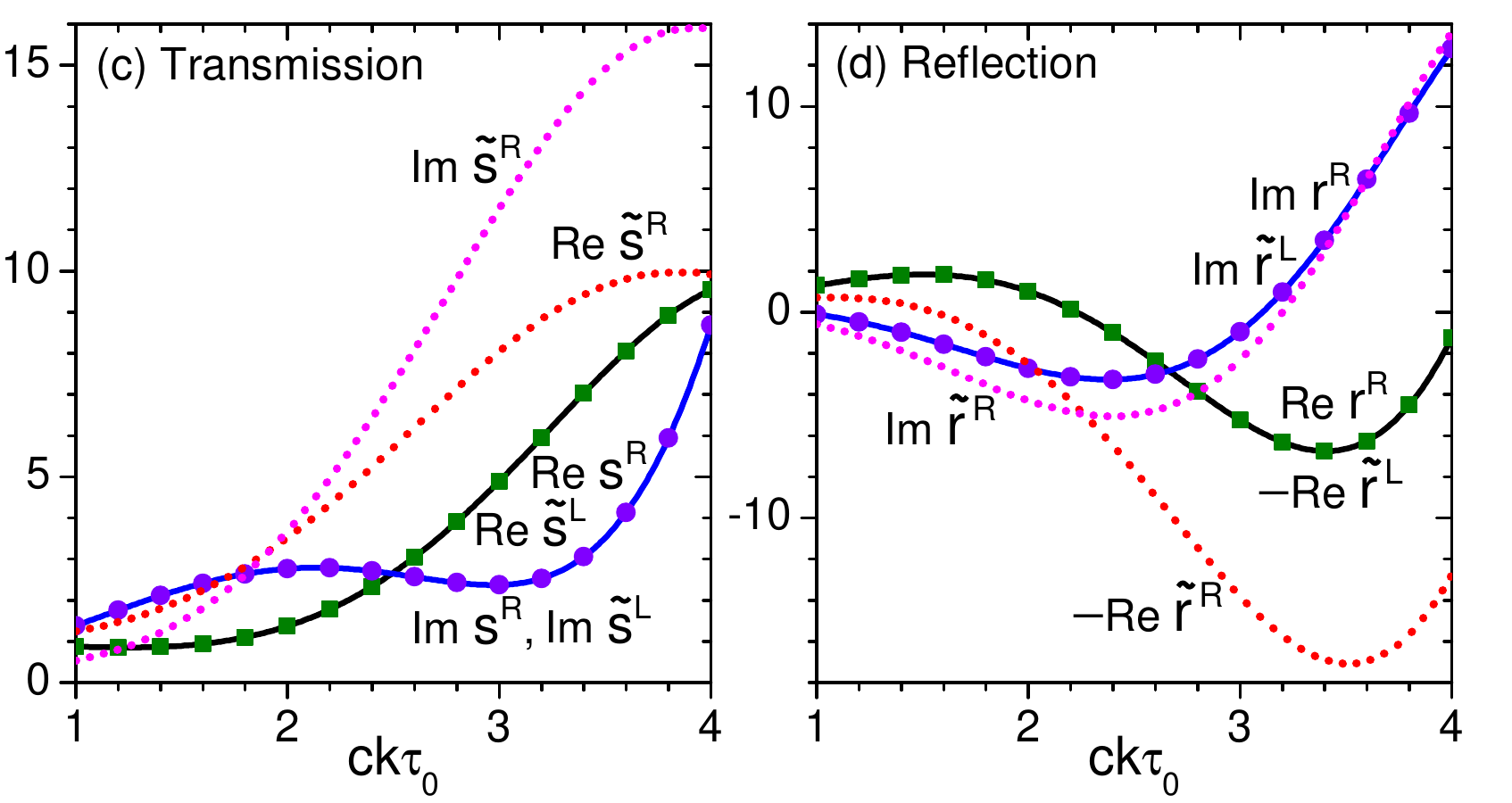}
  \caption{\textbf{Polarization-exchanging temporal reciprocity in Tellegen media.}
\textbf{a,b} Temporal profiles of $\epsilon(t)$, $\mu(t)$, and $\chi(t)$ for the ordered and reversed modulation sequences.
\textbf{c} Transmission amplitudes as functions of $ck\tau_0$. Channel-preserving invariance fails, $s^R \neq \tilde s^R$, but is fully restored under circular-polarization exchange, $s^R=\tilde s^L$, confirming Eq.~(\ref{eq:cesym}).
\textbf{d} Reflection amplitudes exhibit the corresponding symmetry: the single-channel relation $r^R = -(\tilde r^R)^*$ does not hold, whereas the channel-exchanging relation $r^R = -(\tilde r^L)^*$ is satisfied exactly. Real and imaginary parts are shown separately in both panels.}
  \label{fig:num_tellegen}
\end{figure}

\section{Discussion}

We have established a symmetry principle governing temporal scattering
in spatially homogeneous electromagnetic media. By formulating time-dependent
wave evolution at fixed wave vector within a temporal scattering-matrix
framework, we have shown that, when the initial and final media coincide, the
scattering matrices of ordered and reversed modulation sequences are related by
inverse--conjugation, independently of the number of stages and for both
propagating and evanescent intermediate intervals. Temporal reciprocity therefore
arises as a structural property of the temporal scattering algebra, rather than
as a feature of particular modulation protocols.

The classification emerging from this principle is dictated by the
magnetoelectric constitution and is summarized in Table~1. Isotropic and chiral
media form a channel-preserving class: sequence reversal enforces
inverse--conjugation symmetry within each circular-polarization channel, and the
pseudo-unitary \(\mathrm{SU}(1,1)\) structure of lossless temporal scattering is
preserved. Chirality modifies only the accumulated propagation phases, leaving
the temporal interface coefficients and the underlying symmetry algebra
unchanged. Isotropic and chiral media are therefore governed by the same
temporal-reciprocity structure despite their different spatial optical
properties.

Tellegen media constitute a distinct channel-exchanging class. Because their
effective impedances depend on circular polarization, temporal interfaces are
polarization dependent, and Lorentz reciprocity is violated in the spatial
sense. Yet temporal reciprocity survives: sequence reversal must be accompanied
by exchange of circular-polarization channels, equivalently by reversal of the
Tellegen parameter in the interface coefficients. Thus spatial nonreciprocity
and temporal reciprocity refer to different scattering problems. Magnetoelectric
asymmetry reorganizes temporal reciprocity rather than breaking it.

A direct consequence of this classification is deterministic time rewinding, the
exact reconstruction of time-evolving wave states in reversed temporal order.
For each material class, this reconstruction follows from the corresponding
inverse--conjugation relation without requiring parameter-by-parameter retracing
of the modulation history. The explicit rewinding conditions derived here
provide design rules for isotropic, chiral, and magnetoelectric platforms. In
the Tellegen class, rewinding requires simultaneous sign reversal of
\(\epsilon\), \(\mu\), and \(\chi\) in the conjugate medium, a condition that is
physically meaningful in view of the recent experimental demonstration of giant
and tunable Tellegen responses in microwave metamaterials~\cite{gtel}. This
suggests that polarization-exchanging temporal reciprocity and its rewinding
corollary should be experimentally accessible in time-modulated magnetoelectric
systems.

More broadly, the inverse--conjugation symmetry reveals hidden pseudo-unitary
structures in time-modulated wave dynamics. In the isotropic and chiral classes,
temporal scattering is described by a single \(\mathrm{SU}(1,1)\) sector; in the
Tellegen class, it comprises two \(\mathrm{SU}(1,1)\) sectors linked by channel
exchange. This algebraic picture naturally accommodates negative-index and
evanescent intervals and identifies closure of the temporal scattering algebra
under composition as the structural property underlying temporal
reciprocity identified here.

The same viewpoint also applies to continuously time-periodic modulation and
Floquet media. For a finite modulation window \(0<t<T\), let
$f(t)=\{\epsilon(t),\mu(t),\gamma(t),\chi(t)\}$
denote the time-dependent constitutive parameters, with time-reversed
counterpart \(f_{\rm rev}(t)=f(T-t)\). Since a continuous profile is obtained as
the continuum limit of a piecewise temporal sequence, the inverse--conjugation
relation gives $\mathcal S[f(t)]
=\left(\mathcal S[f(T-t)]^*\right)^{-1}$,
with circular-channel exchange in the Tellegen class. For a periodic drive, this
relation applies to a finite time-periodic slab containing one or more periods,
or to the one-period Floquet evolution operator in a lossless system.
Parametric amplification, as in photonic time crystals, is therefore not
incompatible with temporal reciprocity; it is paired with de-amplification in
the inverse-conjugated reversed ordering. In a time-rewinding protocol this
pairing becomes exact,
$\mathcal S_{\rm rev}^{*}\mathcal S_{\rm ord}=I$,
so that the gain accumulated during the ordered evolution is undone by the
conjugate reversed evolution. Recent work on time-periodic chiral media, where
continuous modulation gives rise to handedness-dependent momentum gaps and
parametric amplification, provides a concrete Floquet setting in which this
symmetry perspective may be useful~\cite{kou}.

The fixed-\(k\) formulation also clarifies the role of dispersion. In a
spatially homogeneous dispersive medium, temporal modulation still conserves
\(k\), so different \(k\) components evolve independently. Dispersion changes
the eigenfrequency branches, modal basis, impedances, and temporal matching
coefficients for each \(k\), but it does not by itself destroy the ordering
symmetry. Once the lossless dispersive dynamics is formulated in the appropriate
conservative state space, the inverse--conjugation relation applies separately
to each fixed-\(k\) sector. A finite pulse therefore inherits the same relation
component by component, provided the reversed or conjugate protocol is
implemented over the pulse bandwidth. Material absorption or externally supplied
dissipative gain is different, because it deforms the simple pseudo-unitary
\(\mathrm{SU}(1,1)\) structure of the reduced two-branch scattering matrix. The
exact relations in Eqs.~(\ref{eq:conjrevw}) and
~(\ref{eq:tellegen_reciprocity}), as written, are therefore lossless results. A
generalized treatment of lossy temporal media requires modified scattering
matrices and represents a natural extension of the present framework.

The universality of the inverse--conjugation symmetry, which rests on the
closure properties of the temporal scattering algebra rather than on features
specific to electromagnetic waves, suggests that analogous reciprocity classes
should exist in acoustic, elastic, and quantum wave systems. The identification
of deterministic time rewinding in Dirac-wave systems~\cite{rewind} supports
this expectation and points toward a unified symmetry theory of temporal
reciprocity across wave physics.

\newpage
\setcounter{equation}{0}
\renewcommand{\theequation}{S\arabic{equation}}
\setcounter{figure}{0}
\renewcommand{\thefigure}{S\arabic{figure}}

\title{\centering \LARGE Symmetry classification of temporal reciprocity in
time-varying electromagnetic media: Supplement}

\maketitle

\section*{Note 1. Temporal reciprocity in isotropic media}
\label{sec1}

The elements of a temporal interface scattering matrix are fully determined by the electromagnetic parameters of the two media forming the interface. Throughout this work, the permittivity $\epsilon$ and permeability $\mu$ are assumed real, but not restricted in sign.

When $\epsilon$ and $\mu$ have the same sign, the refractive index
\[
n=\sqrt{\epsilon}\sqrt{\mu}
\]
and the wave impedance
\[
\eta=\frac{\sqrt{\mu}}{\sqrt{\epsilon}}
\]
are real, and the medium supports propagating modes, including both positive-index ($\epsilon>0$, $\mu>0$) and negative-index ($\epsilon<0$, $\mu<0$) cases. When $\epsilon$ and $\mu$ have opposite signs, $n$ and $\eta$ are purely imaginary, and the medium supports only evanescent modes.

If both media forming the interface are either propagating or evanescent, all elements of the temporal interface scattering matrix are real. When one medium is propagating and the other is evanescent, the interface coefficients satisfy
\begin{align}
r_{\mathrm{A}\to\mathrm{B}} = s_{\mathrm{A}\to\mathrm{B}}^{*},
\nonumber
\end{align}
a relation that follows directly from the impedance ratio defined in Eq.~(5).

\subsection*{Propagation matrix}

Temporal evolution within an interval of constant material parameters is described by a propagation matrix $\mathcal P$. In the eigenfrequency basis, this matrix is diagonal and satisfies the general identity
\begin{align}
\mathcal P^{-1} = M \mathcal P M,
\label{eq:perm}
\end{align}
where
\begin{align}
M =
\begin{pmatrix}
0 & 1 \\
1 & 0
\end{pmatrix}
\nonumber
\end{align}
is the permutation matrix that exchanges the two eigenfrequency branches.
Left multiplication by $M$ interchanges matrix rows, while right multiplication interchanges columns, corresponding to exchange of the two frequency components.

If the medium is propagating, the accumulated phase $\phi$ is real and the propagation matrix takes the form
\begin{align}
    \mathcal P=\begin{pmatrix}
        e^{-i\phi} & 0 \\
        0 & e^{i\phi}
    \end{pmatrix}.
    \nonumber
\end{align}
In this case, the matrix additionally satisfies
\begin{align}
\mathcal P^{-1} = \mathcal P^{*},
\nonumber
\end{align}
which is consistent with Eq.~(\ref{eq:perm}).

If the medium is evanescent, the propagation matrix remains diagonal but becomes real,
\begin{align}
    \mathcal P=\begin{pmatrix}
        e^{\kappa} & 0 \\
        0 & e^{-\kappa}
    \end{pmatrix},
    \nonumber
\end{align}
with real decay parameter $\kappa$. In this case,
\begin{align}
\mathcal P^{-1} = M \mathcal P M,
\nonumber
\end{align}
and no additional complex-conjugation relation arises.

\subsection*{Three-stage process}

Consider the three-stage temporal sequence
\[
\mathrm A \to \mathrm B \to \mathrm C,
\]
whose temporal scattering matrix is
\begin{align}
\mathcal S_{\mathrm{A}\to\mathrm{B}\to\mathrm{C}} =
\mathcal S_{\mathrm{B}\to\mathrm{C}}
\mathcal P_{\mathrm B}
\mathcal S_{\mathrm{A}\to\mathrm{B}}.
\label{eq:SABC}
\end{align}
Here $\mathcal S_{i\to j}$ denotes the temporal interface matrix from medium $i$ to $j$, and $\mathcal P_{\mathrm B}$ is the propagation matrix of the intermediate slab.
Temporal interface matrices obey two fundamental algebraic identities:

(i) Composition of successive interfaces
\begin{align}
\mathcal S_{\mathrm{A}\to\mathrm{C}} =
\mathcal S_{\mathrm{B}\to\mathrm{C}}
\mathcal S_{\mathrm{A}\to\mathrm{B}},
\label{eq:comp1}
\end{align}

(ii) Reversal as inversion
\begin{align}
\mathcal S_{\mathrm{A}\to\mathrm{B}} =
\mathcal S_{\mathrm{B}\to\mathrm{A}}^{-1}.
\label{eq:inv}
\end{align}
As a consistency check, if the intermediate slab is infinitesimally short ($\phi_{\mathrm B}=0$, so that $\mathcal P_{\mathrm B}=I$), Eq.~(\ref{eq:SABC}) reduces to the direct interface relation
\[
\mathcal S_{\mathrm{A}\to\mathrm{B}\to\mathrm{C}}
=
\mathcal S_{\mathrm{A}\to\mathrm{C}},
\]
in agreement with Eq.~(\ref{eq:comp1}).

We now assume that the initial and final media, $\mathrm A$ and $\mathrm C$, are propagating. Using the inversion property in Eq.~(\ref{eq:inv}), Eq.~(\ref{eq:SABC}) can be rewritten as
\begin{align}
\mathcal S_{\mathrm{A}\to\mathrm{B}\to\mathrm{C}}
=
\left(
\mathcal S_{\mathrm{B}\to\mathrm{A}}
\mathcal P_{\mathrm B}^{-1}
\mathcal S_{\mathrm{C}\to\mathrm{B}}
\right)^{-1}.
\label{eq:rewrite}
\end{align}
Equation~(\ref{eq:rewrite}) expresses the ordered sequence in terms of the reversed sequence and forms the starting point for establishing the inverse–conjugation symmetry.

\paragraph*{Propagating intermediate medium}

If the intermediate slab $\mathrm B$ is propagating,
the accumulated phase is real and the propagation matrix satisfies
\begin{align}
  \mathcal P_{\mathrm B}^{-1}=\mathcal P_{\mathrm B}^{*}.
  \nonumber
\end{align}
Since the interface matrices are real when both adjacent media are of the same type (propagating in the present case), Eq.~(\ref{eq:rewrite}) immediately gives
\begin{align}
\mathcal S_{\mathrm{A}\to\mathrm{B}\to\mathrm{C}}
=
\left[
\left(
\mathcal S_{\mathrm{B}\to\mathrm{A}}
\mathcal P_{\mathrm B}
\mathcal S_{\mathrm{C}\to\mathrm{B}}
\right)^{*}
\right]^{-1}.
\label{eq:prop_case}
\end{align}

\paragraph*{Evanescent intermediate medium}

If $\mathrm B$ is evanescent, the propagation matrix is real and diagonal. Its inverse is obtained through the permutation identity
\[
\mathcal P_{\mathrm B}^{-1}
=
M \mathcal P_{\mathrm B} M.
\]
In addition, the interface matrices satisfy
\begin{align}
\mathcal S_{i\to j}^{*}
=
M \mathcal S_{i\to j}
=
\mathcal S_{i\to j} M,
\label{eq:intrm}
\end{align}
which follows from the purely imaginary impedance ratio in the propagating–evanescent case.
Substituting these relations into Eq.~(\ref{eq:rewrite}) and commuting the permutation matrices through the product, one finds that all permutation operations cancel pairwise. The resulting expression reduces again to
\begin{align}
\mathcal S_{\mathrm{A}\to\mathrm{B}\to\mathrm{C}}
=
\left[
\left(
\mathcal S_{\mathrm{B}\to\mathrm{A}}
\mathcal P_{\mathrm B}
\mathcal S_{\mathrm{C}\to\mathrm{B}}
\right)^{*}
\right]^{-1}.
\label{eq:ev_case}
\end{align}

\paragraph*{Unified result}

Equations~(\ref{eq:prop_case}) and~(\ref{eq:ev_case}) are identical. Therefore, irrespective of whether the intermediate slab is propagating or evanescent, the three-stage process satisfies the inverse–conjugation symmetry
\begin{align}
\mathcal S_{\mathrm{A}\to\mathrm{B}\to\mathrm{C}}
=
\left(
\mathcal S_{\mathrm{C}\to\mathrm{B}\to\mathrm{A}}^{*}
\right)^{-1}.
\label{eq:tsic}
\end{align}
The result is thus independent of the nature of the intermediate interval and depends only on the algebraic structure of temporal interfaces and propagation matrices.

\subsection*{Closure of the algebraic structure}

Because the media are lossless and spatially homogeneous, temporal scattering at fixed wavevector conserves the wave-action flux. In the eigenfrequency basis, this conservation law imposes the pseudo-unitarity condition
\begin{align}
\mathcal S^\dagger \Sigma \mathcal S = \Sigma,
\quad
\Sigma =
\begin{pmatrix}
1 & 0 \\
0 & -1
\end{pmatrix}.
\label{eq:pseudo_unitarity}
\end{align}
Equation~(\ref{eq:pseudo_unitarity}) implies that $\mathcal S \in U(1,1)$.
Up to an overall phase factor $e^{i\theta}$, the most general $2\times2$ matrix satisfying this condition can be written as
\begin{align}
\mathcal S
=
e^{i\theta}
\begin{pmatrix}
s & r^{*} \\
r & s^{*}
\end{pmatrix},
\quad
|s|^{2}-|r|^{2}=1.
\nonumber
\end{align}
Since an overall phase does not affect temporal transmission or reflection coefficients, we fix the phase convention such that
\begin{align}
\mathcal S =
\begin{pmatrix}
s & r^{*} \\
r & s^{*}
\end{pmatrix}.
\label{eq:universal_form1}
\end{align}
This canonical form is characteristic of lossless temporal scattering in isotropic media and is preserved under successive temporal modulations.

To verify that this algebraic structure is closed under successive temporal modulations, consider two matrices of the form~(\ref{eq:universal_form1}),
\begin{align}
\mathcal S_1 =
\begin{pmatrix}
s_1 & r_1^{*} \\
r_1 & s_1^{*}
\end{pmatrix},
\quad
\mathcal S_2 =
\begin{pmatrix}
s_2 & r_2^{*} \\
r_2 & s_2^{*}
\end{pmatrix}.
\nonumber
\end{align}
Their product is
\begin{align}
\mathcal S_2 \mathcal S_1
=
\begin{pmatrix}
s_2 s_1 + r_2^{*} r_1 &
s_2 r_1^{*} + r_2^{*} s_1^{*} \\
r_2 s_1 + s_2^{*} r_1 &
r_2 r_1^{*} + s_2^{*} s_1^{*}
\end{pmatrix}.
\nonumber
\end{align}
Defining
\begin{align}
s_3 = s_2 s_1 + r_2^{*} r_1,
\quad
r_3 = r_2 s_1 + s_2^{*} r_1,
\nonumber
\end{align}
the product can be written as
\begin{align}
\mathcal S_2 \mathcal S_1 =
\begin{pmatrix}
s_3 & r_3^{*} \\
r_3 & s_3^{*}
\end{pmatrix},
\nonumber
\end{align}
which is again of the form~(\ref{eq:universal_form1}).
Hence the set of temporal scattering matrices satisfying the flux-conservation constraint is closed under multiplication. Consequently, arbitrarily long temporal modulation sequences preserve the same algebraic structure, ensuring that the inverse–conjugation symmetry derived for a three-stage process extends to general multi-stage modulations.

\subsection*{Extension to arbitrary sequences}

We now extend the inverse–conjugation symmetry from a three-stage process to an arbitrary temporal modulation sequence.
Consider an ordered sequence of $N$ temporal slabs with identical initial and final (propagating) media $\mathrm I$,
\begin{align}
\mathcal S_{\mathrm{ord}}
=
\mathcal S_{N\to\mathrm I}
\mathcal P_N
\mathcal S_{N-1\to N}
\mathcal P_{N-1}
\cdots
\mathcal S_{1\to2}
\mathcal P_1
\mathcal S_{\mathrm I\to1}.
\nonumber
\end{align}
Using the interface-composition identity ($\mathcal{S}_{j\to j+1}=\mathcal{S}_{\mathrm{I}\to j+1} \mathcal{S}_{j\to \mathrm{I}}$), each interface can be expressed relative to the reference medium
$\mathrm I$. The ordered sequence can then be regrouped into three-stage blocks of the form
\[
Q_j\equiv \mathcal S_{j\to\mathrm I}
\mathcal P_j
\mathcal S_{\mathrm I\to j},
\]
so that
\begin{align}
    \mathcal S_{\mathrm{ord}}
&=Q_NQ_{N-1}\cdots Q_2Q_1\nonumber.
\end{align}
As shown in Eq.~(\ref{eq:tsic}), each block satisfies the inverse–conjugation symmetry
\begin{align}
    Q_j=\left(Q_j^*\right)^{-1},
    \nonumber
\end{align}
independently of whether slab $j$ is propagating or evanescent. Applying this identity successively and using the reversal of matrix order under inversion, we obtain
\begin{align}
    \mathcal S_{\mathrm{ord}}
&=Q_NQ_{N-1}\cdots Q_2Q_1\nonumber\\
&=\left[\left(Q_1Q_{2}\cdots Q_{N-1}Q_N\right)^*\right]^{-1}.
\nonumber
\end{align}
The product inside the parentheses is exactly the scattering matrix associated with the reversed temporal modulation sequence,
\begin{align}
    \mathcal S_{\mathrm{rev}}=Q_1Q_2\cdots Q_N.
    \nonumber
\end{align}
Hence,
\begin{align}
\mathcal S_{\mathrm{ord}}
=
\left(
\mathcal S_{\mathrm{rev}}^{*}
\right)^{-1}.
\label{eq:rewinding_iso}
\end{align}
This establishes that the inverse–conjugation symmetry holds for arbitrary temporal modulation sequences, independent of the number of slabs and of the propagating or evanescent character of intermediate intervals.

\section*{Note 2. Time-rewinding conditions in isotropic media from temporal reciprocity}
\label{sec:s2}

We assume that the initial and final media are identical and denoted by $\mathrm A$.
Consider the ordered temporal sequence
\[
\mathrm A \to \mathrm B \to \mathrm C \to \mathrm A,
\]
together with its reversed counterpart
\[
\mathrm A \to \mathrm C \to \mathrm B \to \mathrm A.
\]
As established in Note 1, the associated scattering matrices
\[
\mathcal S_{\mathrm{ord}}
=
\mathcal S_{\mathrm{A}\to\mathrm{B}\to\mathrm{C}\to\mathrm{A}},
\quad
\mathcal S_{\mathrm{rev}}
=
\mathcal S_{\mathrm{A}\to\mathrm{C}\to\mathrm{B}\to\mathrm{A}}
\]
satisfy the temporal reciprocity identity
\begin{align}
\mathcal S_{\mathrm{rev}}^{*}\,\mathcal S_{\mathrm{ord}} = I,
\nonumber
\end{align}
for arbitrary isotropic intermediate media $\mathrm B$ and $\mathrm C$.

The ordered sequence admits the factorization
\begin{align}
\mathcal S_{\mathrm{ord}} = Q_{\mathrm C}\,Q_{\mathrm B},
\nonumber
\end{align}
where
\begin{align}
Q_{\mathrm B}
  = \mathcal S_{\mathrm{B}\to\mathrm A}\,
     \mathcal P_{\mathrm B}\,
     \mathcal S_{\mathrm A\to\mathrm B}, \quad
Q_{\mathrm C}
  = \mathcal S_{\mathrm{C}\to\mathrm A}\,
     \mathcal P_{\mathrm C}\,
     \mathcal S_{\mathrm A\to\mathrm C}.
     \nonumber
\end{align}
Similarly, the reversed sequence factorizes as
\begin{align}
\mathcal S_{\mathrm{rev}} = Q_{\mathrm B}\,Q_{\mathrm C}.
\nonumber
\end{align}
Substituting these into the reciprocity identity yields
\begin{align}
Q_{\mathrm B}^{*}\,Q_{\mathrm C}^{*}\,Q_{\mathrm C}\,Q_{\mathrm B} = I.
\nonumber
\end{align}

\begin{figure}
  \centering
  \includegraphics[width=9cm]{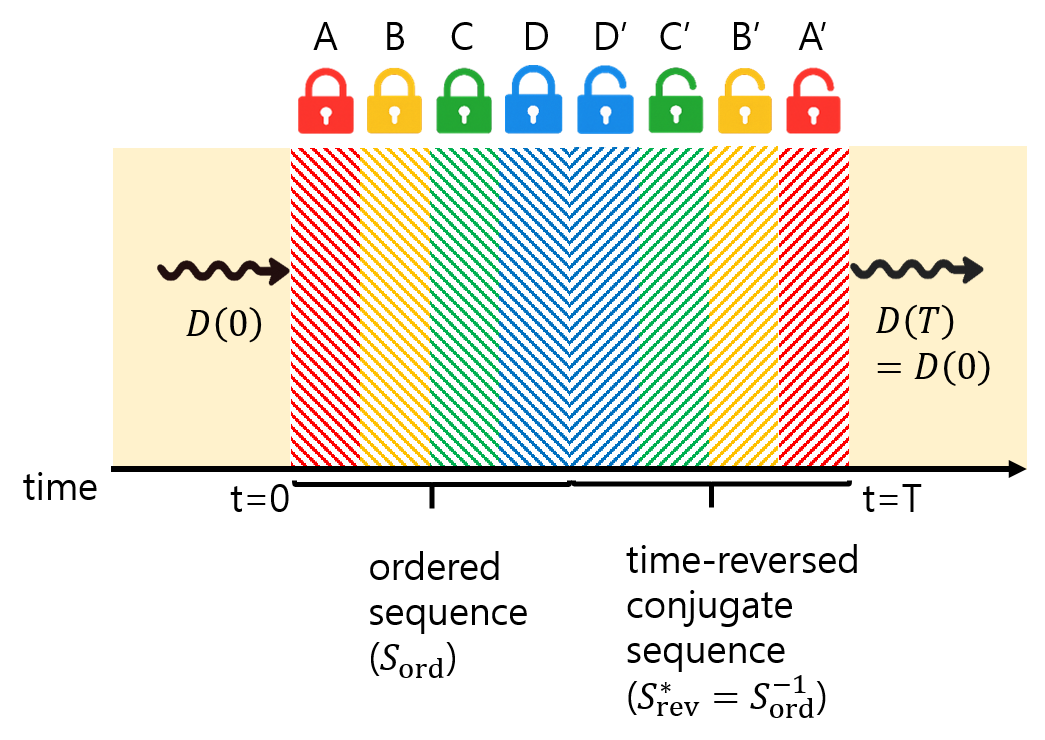}
  \caption{Time-rewinding based on temporal reciprocity.
An incident wave first undergoes scattering by an ordered temporal modulation sequence ${\mathrm A}\to{\mathrm B}\to{\mathrm C}\to{\mathrm D}$, producing transmitted and reflected components determined by the corresponding temporal scattering matrix. The subsequent application of the time-reversed conjugate sequence ${\mathrm D'}\to{\mathrm C'}\to{\mathrm B'}\to{\mathrm A'}$ implements the inverse evolution. As a result, the combined modulation satisfies $\mathcal S_{\mathrm{rev}}^{*}\mathcal S_{\mathrm{ord}}=I$ and restores the field to its initial state, $D(T)=D(0)$. The overall process thus reverses the temporal evolution of the wave, realizing deterministic time rewinding as a direct consequence of temporal reciprocity.}
\label{fig2}
\end{figure}

We now construct a temporal modulation designed to achieve deterministic time rewinding:
\[
\mathrm A \to \mathrm B \to \mathrm C
              \to \mathrm C' \to \mathrm B' \to \mathrm A,
\]
where $(\mathrm B,\mathrm B')$ and $(\mathrm C,\mathrm C')$
form conjugate pairs. The corresponding scattering matrix is
\begin{align}
\mathcal S = Q_{\mathrm B'}\,Q_{\mathrm C'}\,Q_{\mathrm C}\,Q_{\mathrm B}.
\nonumber
\end{align}
Perfect time rewinding, $\mathcal S=I$, is achieved provided that
\begin{align}
Q_{\mathrm B'} = Q_{\mathrm B}^{*},
\quad
Q_{\mathrm C'} = Q_{\mathrm C}^{*}.
\nonumber
\end{align}
Under these conditions, the second half of the sequence exactly inverts the first, restoring the initial wave state for arbitrary input amplitudes and phases.
A schematic illustration of this reciprocity-based time-rewinding protocol is shown in Fig.~\ref{fig2}.

\subsection*{Propagating conjugate media}

We first assume that both $\mathrm B$ and $\mathrm B'$ are propagating media.
For a three-stage process
$\mathrm A \to \mathrm B \to \mathrm A$
with propagating intermediate medium $\mathrm B$, the scattering matrix reads
\begin{align}
Q_{\mathrm B}
&=\mathcal{S}_{\mathrm{A}\to \mathrm{B} \to \mathrm{A}}
\nonumber\\
&=
\begin{pmatrix}
s_{\mathrm{B}\to\mathrm A}s_{\mathrm A\to\mathrm B}
&
s_{\mathrm{B}\to\mathrm A}r_{\mathrm A\to\mathrm B} \\
r_{\mathrm{B}\to\mathrm A}s_{\mathrm A\to\mathrm B}
&
r_{\mathrm{B}\to\mathrm A}r_{\mathrm A\to\mathrm B}
\end{pmatrix}
e^{-i\phi_{\mathrm B}}
+
\begin{pmatrix}
r_{\mathrm{B}\to\mathrm A}r_{\mathrm A\to\mathrm B}
&
r_{\mathrm{B}\to\mathrm A}s_{\mathrm A\to\mathrm B} \\
s_{\mathrm{B}\to\mathrm A}r_{\mathrm A\to\mathrm B}
&
s_{\mathrm{B}\to\mathrm A}s_{\mathrm A\to\mathrm B}
\end{pmatrix}
e^{i\phi_{\mathrm B}},
\nonumber
\end{align}
where $\phi_{\mathrm B}=\omega_{\mathrm B}\tau_{\mathrm B}$.
For propagating media, the interface coefficients are real and given by
\begin{align}
s_{\mathrm A\to\mathrm B}
= \tfrac12\!\left(1+\frac{\eta_{\mathrm A}}{\eta_{\mathrm B}}\right),
\quad
r_{\mathrm A\to\mathrm B}
= \tfrac12\!\left(1-\frac{\eta_{\mathrm A}}{\eta_{\mathrm B}}\right).
\nonumber
\end{align}
The time-rewinding condition
\begin{align}
Q_{\mathrm B'} = Q_{\mathrm B}^{*}
\nonumber
\end{align}
is satisfied if and only if
\begin{align}
\phi_{\mathrm B'} = -\phi_{\mathrm B},
\quad
\eta_{\mathrm B'} = \eta_{\mathrm B}.
\label{eq:rewind_cond_prop}
\end{align}
Using $\phi_{\mathrm B}=\omega_{\mathrm B}\tau_{\mathrm B}$ with
$\omega_{\mathrm B}=ck/n_{\mathrm B}$,
conditions~\eqref{eq:rewind_cond_prop} imply
\begin{align}
\frac{\epsilon_{\mathrm B'}}{\epsilon_{\mathrm B}}
=
\frac{\mu_{\mathrm B'}}{\mu_{\mathrm B}},
\quad
\frac{\tau_{\mathrm B'}}{\tau_{\mathrm B}}
=
\left|
\frac{n_{\mathrm B'}}{n_{\mathrm B}}
\right|,
\label{eq:trca22}
\end{align}
where
\[
n_{\mathrm B}=\sqrt{\epsilon_{\mathrm B}}\sqrt{\mu_{\mathrm B}}.
\]
Moreover,
\begin{align}
\operatorname{sgn}\!\left(\frac{\epsilon_{\mathrm B'}}{\epsilon_{\mathrm B}}\right)
=
\operatorname{sgn}\!\left(\frac{\mu_{\mathrm B'}}{\mu_{\mathrm B}}\right)
=-1,
\label{eq:trca23w}
\end{align}
so that the conjugate media have opposite refractive indices but identical impedances.
The slab durations are not required to be equal; the ratio $\tau_{\mathrm B'}/\tau_{\mathrm B}$ may take any positive value consistent with Eq.~\eqref{eq:trca22}. This regime reproduces the total-transmission–based time rewinding analyzed in Ref.~[12].

\subsection*{Evanescent conjugate media}

If $\mathrm B$ and $\mathrm B'$ are evanescent media,
the three-stage scattering matrix takes the form
\begin{align}
Q_{\mathrm B}
&=
\begin{pmatrix}
s_{\mathrm{B}\to\mathrm A}s_{\mathrm{A}\to \mathrm B} &
s_{\mathrm{B}\to\mathrm A}r_{\mathrm{A}\to \mathrm B} \\
r_{\mathrm{B}\to\mathrm A}s_{\mathrm{A}\to \mathrm B} &
r_{\mathrm{B}\to\mathrm A}r_{\mathrm{A}\to \mathrm B}
\end{pmatrix}
e^{\kappa_{\mathrm B}}
+
\begin{pmatrix}
r_{\mathrm{B}\to\mathrm A}r_{\mathrm{A}\to \mathrm B} &
r_{\mathrm{B}\to\mathrm A}s_{\mathrm{A}\to \mathrm B} \\
s_{\mathrm{B}\to\mathrm A}r_{\mathrm{A}\to \mathrm B} &
s_{\mathrm{B}\to\mathrm A}s_{\mathrm{A}\to \mathrm B}
\end{pmatrix}
e^{-\kappa_{\mathrm B}},
\nonumber
\end{align}
where $\kappa_{\mathrm B}>0$ is the decay parameter.
In this regime the interface coefficients are complex and obey
\[
s_{\mathrm{B}\to\mathrm A}^{*}
=
r_{\mathrm{B}\to\mathrm A},
\]
a direct consequence of the imaginary impedance ratio when one of the two media is evanescent.
The time-rewinding condition
\begin{align}
Q_{\mathrm B'} = Q_{\mathrm B}^{*}
\nonumber
\end{align}
is satisfied if and only if
\begin{align}
\kappa_{\mathrm B'} = \kappa_{\mathrm B},
\qquad
\eta_{\mathrm B'} = -\eta_{\mathrm B}.
\label{eq:rewind_cond_evan}
\end{align}
Since
\[
\kappa_{\mathrm B}
=
\frac{ck\,\tau_{\mathrm B}}{|n_{\mathrm B}|},
\]
conditions~\eqref{eq:rewind_cond_evan} reduce to
Eqs.~\eqref{eq:trca22} and \eqref{eq:trca23w}, with $n_{\mathrm B}$ purely imaginary. In this case, the conjugate evanescent media exhibit identical decay rates while having opposite impedances. As in the propagating regime, the slab durations need not coincide; only the ratio required by the decay condition must be satisfied. This configuration reproduces the total-reflection--based time-rewinding mechanism discussed in Ref.~[12].

\section*{Note 3. Circular eigenmodes and dispersion in bi-isotropic media}

\subsection*{Bi-isotropic wave equations}

We begin with the bi-isotropic constitutive relations
\[
\mathbf D=\epsilon \mathbf E + a \mathbf H,
\quad
\mathbf B=\mu \mathbf H + a^{*}\mathbf E,
\]
where the complex magnetoelectric parameter is
\begin{align}
    a=\chi+i\gamma,
    \nonumber
\end{align}
with $\chi$ the Tellegen parameter and $\gamma$ the chirality parameter.
In a spatially homogeneous medium, Maxwell’s equations yield coupled wave equations for the transverse field components. For plane waves with wavevector $\mathbf k = k\hat{\mathbf z}$, the system decouples in the circular basis
\[
D_L=\frac{D_x+iD_y}{\sqrt2},
\quad
D_R=\frac{D_x-iD_y}{\sqrt2},
\]
which diagonalizes the magnetoelectric coupling.
The resulting time-domain equations for the two independent channels are
\begin{align}
\frac{d}{dt}\!\left(\frac{\epsilon\mu-|a|^2}{\epsilon}\frac{dD_L}{dt}
-\frac{cka^{*}}{\epsilon}D_L\right)
+\frac{cka}{\epsilon}\frac{dD_L}{dt}
+\frac{c^2k^2}{\epsilon}D_L &=0,
\nonumber\\
\frac{d}{dt}\!\left(\frac{\epsilon\mu-|a|^2}{\epsilon}\frac{dD_R}{dt}
+\frac{cka^{*}}{\epsilon}D_R\right)
-\frac{cka}{\epsilon}\frac{dD_R}{dt}
+\frac{c^2k^2}{\epsilon}D_R &=0.
\label{eq:bimm}
\end{align}
Thus, at fixed $k$, the $D_L$ and $D_R$ channels are dynamically decoupled. Temporal scattering can therefore be treated independently in each circular channel.

\subsection*{Dispersion relations}

In a temporally stationary interval, substituting the harmonic ansatz
\begin{align}
    D_{L,R}\propto e^{-i\omega t}
    \nonumber
\end{align}
into Eq.~\eqref{eq:bimm} yields a quadratic dispersion relation in each channel. For the $D_L$ channel,
\begin{align}
c^2k^2-\omega^2(\epsilon\mu-\chi^2-\gamma^2)
+2\omega ck\gamma=0.
\nonumber
\end{align}
Solving for $\omega$ gives
\begin{align}
\omega_{\pm}^{[L]}(k)
=
\frac{ck}{\epsilon\mu-\chi^2-\gamma^2}
\left(
\gamma \pm \sqrt{\epsilon\mu-\chi^2}
\right).
\label{eq:dispbil}
\end{align}
The superscript $[L]$ indicates that the solution arises from the $D_L$ eigenproblem; it does not specify propagation direction. For fixed $k>0$, the sign of $\omega$ determines the direction of energy transport. The two branches therefore correspond to opposite temporal phase evolution and, under the conventions adopted in this work, to opposite propagation directions.

Repeating the same procedure for the $D_R$ channel yields
\begin{align}
\omega_{\pm}^{[R]}(k)
=
\frac{ck}{\epsilon\mu-\chi^2-\gamma^2}
\left(
-\gamma \pm \sqrt{\epsilon\mu-\chi^2}
\right).
\label{eq:dispbir}
\end{align}
In general, these branches are not related by simple sign inversion ($\omega \neq \pm \omega_0$); instead they reflect the intrinsic magnetoelectric coupling of the medium.

A useful symmetry relation is
\begin{align}
\omega_{+}^{[R]}=-\omega_{-}^{[L]},
\quad
\omega_{-}^{[R]}=-\omega_{+}^{[L]}.
\nonumber
\end{align}
Thus, the negative-frequency branch in one circular channel corresponds to the positive-frequency branch in the opposite channel. In particular, a reflected wave appearing in the $D_R$ channel with negative temporal phase evolution corresponds to left-circular polarization (LCP), while the analogous branch in the $D_L$ channel corresponds to right-circular polarization (RCP), consistent with the polarization conventions adopted in the main text.

\subsection*{Temporal transmission and reflection at a temporal interface}

Consider a temporal interface at $t=0$ between media $\mathrm A$ and $\mathrm B$.
From Eq.~\eqref{eq:bimm}, the quantities that must remain continuous across the interface are
\begin{align}
    D_{L,R} \quad \frac{\epsilon\mu-|a|^2}{\epsilon}\frac{dD_{L,R}}{dt}
\mp\frac{cka^{*}}{\epsilon}D_{L,R},
\nonumber
\end{align}
where the upper (lower) sign applies to the $D_L$ ($D_R$) channel.
For example, in the $D_R$ channel, assume an incident wave propagating in the $+z$ direction:
\begin{align}
D_R=
\begin{cases}
e^{-i\omega_{{\mathrm {A}}+}^{[R]}t}, & t<0,\\
s_{\mathrm {A}\to{\mathrm B}}^Re^{-i\omega_{{\mathrm {B}}+}^{[R]}t}+r_{\mathrm {A}\to{\mathrm B}}^Re^{-i\omega_{{\mathrm {B}}-}^{[R]}t}, & t>0.
\end{cases}
\nonumber
\end{align}
Continuity of $D_R$ gives
\begin{align}
    s_{\mathrm {A}\to{\mathrm B}}^R+r_{\mathrm {A}\to{\mathrm B}}^R=1.
    \nonumber
\end{align}
Applying continuity of the relevant derivative combination across the temporal interface and performing straightforward algebra, we obtain
\begin{align}
s_{\mathrm {A}\to{\mathrm B}}^{R}=\tfrac12\left(1+\rho_{\mathrm {A}\to{\mathrm B}}- i\nu_{\mathrm {A}\to{\mathrm B}}\right),\quad r_{\mathrm {A}\to{\mathrm B}}^{R}=\tfrac12\left(1-\rho_{\mathrm {A}\to{\mathrm B}}+ i\nu_{\mathrm {A}\to{\mathrm B}}\right),
\label{eq:contcond1}
\end{align}
where
\begin{align}
\rho_{\mathrm {A}\to{\mathrm B}}=\frac{\epsilon_{\mathrm B}n_{0\mathrm A}}{\epsilon_{\mathrm A}n_{0\mathrm B}},\quad \nu_{\mathrm {A}\to{\mathrm B}}=\frac{\epsilon_{\mathrm A}\chi_{\mathrm B}-\epsilon_{\mathrm B}\chi_{\mathrm A}}{\epsilon_{\mathrm A}n_{0\mathrm B}},
\nonumber
\end{align}
and
\begin{align}
n_0=
\begin{cases}
\pm\sqrt{\epsilon\mu-\chi^2}, & \epsilon\mu>\chi^2,\\[4pt]
i\sqrt{\chi^2-\epsilon\mu}, & \epsilon\mu<\chi^2,
\end{cases}
\nonumber
\end{align}
with the sign selecting the positive- or negative-index branch.
Repeating the procedure for the $D_L$ channel yields
\begin{align}
s_{\mathrm {A}\to{\mathrm B}}^{L}=\tfrac12\left(1+\rho_{\mathrm {A}\to{\mathrm B}}+ i\nu_{\mathrm {A}\to{\mathrm B}}\right),\quad
r_{\mathrm {A}\to{\mathrm B}}^{L}=\tfrac12\left(1-\rho_{\mathrm {A}\to{\mathrm B}}- i\nu_{\mathrm {A}\to{\mathrm B}}\right).
\label{eq:contcond2}
\end{align}
Equations \eqref{eq:contcond1}–\eqref{eq:contcond2} show explicitly that the Tellegen parameter $\chi$ enters the interface coefficients, whereas chirality $\gamma$ does not affect the temporal matching conditions.
The corresponding interface scattering matrices in the circular basis are therefore
\begin{align}
\mathcal{S}_{\mathrm {A}\to{\mathrm B}}^R=\begin{pmatrix}
  s_{\mathrm {A}\to{\mathrm B}}^R & r_{\mathrm {A}\to{\mathrm B}}^L \\
  r_{\mathrm {A}\to{\mathrm B}}^R & s_{\mathrm {A}\to{\mathrm B}}^L\end{pmatrix},\quad
  \mathcal{S}_{\mathrm {A}\to{\mathrm B}}^L=\begin{pmatrix}
  s_{\mathrm {A}\to{\mathrm B}}^L & r_{\mathrm {A}\to{\mathrm B}}^R \\
  r_{\mathrm {A}\to{\mathrm B}}^L & s_{\mathrm {A}\to{\mathrm B}}^R
\end{pmatrix}.
\nonumber
\end{align}

\subsection*{Polarization-dependent refractive indices and wave impedances}

From Eqs.~\eqref{eq:dispbil}–\eqref{eq:dispbir}, the positive-frequency branches can be written as
\begin{align}
\omega_{+}^{[R]}=\frac{ck}{n_0+\gamma},\quad
\omega_{+}^{[L]}=\frac{ck}{n_0-\gamma}.
\nonumber
\end{align}
The effective refractive indices for right- and left-circular polarization are therefore
\begin{align}
    n_R=n_0+\gamma,\quad n_L=n_0-\gamma.
    \nonumber
\end{align}
Applying the zero-reflection condition via Eqs.~\eqref{eq:contcond1}–\eqref{eq:contcond2} yields the polarization-dependent effective impedances
\begin{align}
    \eta_R=\frac{n_0+i\chi}{\epsilon},\quad \eta_L=\frac{n_0-i\chi}{\epsilon}.
    \nonumber
\end{align}
Importantly, the effective impedances depend on the Tellegen parameter $\chi$ but are independent of the chirality parameter $\gamma$, whereas the refractive indices depend on both.

\section*{Note 4. Temporal reciprocity in chiral media}

In chiral media ($\chi=0$, $\gamma\neq0$), the temporal interface scattering matrix is identical to that of isotropic media. This follows because the temporal matching conditions depend only on the impedance ratio and are independent of the chirality parameter $\gamma$. Chirality therefore does not affect temporal interface scattering.
Its influence appears exclusively in temporal propagation, where it produces an additional phase associated with optical activity. To make this distinction explicit, we examine the structure of the propagation matrix.

Consider a pair of circularly polarized eigenmodes at fixed $k>0$: a forward-propagating RCP mode and a backward-propagating LCP mode. Their angular frequencies are
\begin{align}
\omega_{+}
=
\frac{ck}{n_R}
=
\frac{ck}{n+\gamma},
\quad
\omega_{-}
=
-\frac{ck}{n_L}
=
-\frac{ck}{n-\gamma}.
\nonumber
\end{align}
where $n=\sqrt{\epsilon}\sqrt{\mu}$.
These expressions may be rewritten as
\begin{align}
    \omega_\pm=\frac{ck}{n^2-\gamma^2}\left(\pm n-\gamma\right).
    \nonumber
\end{align}
Each frequency therefore contains two contributions: a term proportional to
\begin{align}
    \frac{ckn}{n^2-\gamma^2},
    \nonumber
\end{align}
which inherits the character of $n$ (real for propagating modes, purely imaginary for evanescent modes),
and a chirality-induced term proportional to
\begin{align}
    \frac{ck\gamma}{n^2-\gamma^2},
    \nonumber
\end{align}
which is strictly real.
Thus, chirality introduces a real frequency shift that enters with the same sign structure in both circular channels. This contribution accounts for optical activity but does not alter the algebraic structure of temporal scattering.

%\subsection*{Inverse–conjugation symmetry}

When the intermediate medium $\mathrm B$ supports propagating modes, both eigenfrequencies are real and the propagation matrix satisfies
\begin{align}
\mathcal P_{\mathrm B}^{-1}=\mathcal P_{\mathrm B}^{*}.
\nonumber
\end{align}
Since the interface matrices are identical to those of isotropic media, the derivation of the inverse–conjugation relation proceeds unchanged:
\begin{align}
\mathcal S_{\mathrm A\to\mathrm B\to\mathrm C}
=
\left(
\mathcal S_{\mathrm C\to\mathrm B\to\mathrm A}^{*}
\right)^{-1}.
\nonumber
\end{align}

If $\mathrm B$ is evanescent, the refractive index $n$ is purely imaginary and the propagation matrix takes the form
\begin{align}
    \mathcal P_{\mathrm B}=\begin{pmatrix}
    e^{\tilde\kappa_{\mathrm{B}}}e^{i\beta_{\mathrm{B}}} & 0 \\
    0 & e^{-\tilde\kappa_{\mathrm{B}}}e^{i\beta_{\mathrm{B}}}
    \end{pmatrix},
    \nonumber
\end{align}
where
\begin{align}
  \tilde\kappa_{\mathrm{B}}=\frac{ck|n_{\mathrm B}|\tau_{\mathrm B}}{n_{\mathrm B}^2-\gamma_{\mathrm B}^2},
  \quad \beta_{\mathrm{B}}=\frac{ck\gamma_{\mathrm B}\tau_{\mathrm B}}{n_{\mathrm B}^2-\gamma_{\mathrm B}^2}.
  \nonumber
\end{align}
Here $\tilde\kappa_{\mathrm B}$ characterizes the exponential attenuation associated with evanescent decay, whereas $\beta_{\mathrm B}$ represents the chirality-induced propagation phase, which remains purely real.
From the diagonal structure of $\mathcal P_{\mathrm B}$ it follows directly that
\begin{align}
\mathcal P_{\mathrm B}^{-1}
=
M\,\mathcal P_{\mathrm B}^{*}\,M,
\nonumber
\end{align}
where $M$ exchanges the two frequency branches. Combining this identity with the interface relations [Eq.~(\ref{eq:intrm})] yields
\begin{align}
\mathcal S_{\mathrm A\to\mathrm B\to\mathrm C}
=
\left(
\mathcal S_{\mathrm C\to\mathrm B\to\mathrm A}^{*}
\right)^{-1}.
\nonumber
\end{align}
Therefore, the inverse–conjugation symmetry of temporal scattering persists in chiral media, independent of whether the intermediate slab supports propagating or evanescent modes. The extension to an arbitrary temporal modulation sequence follows identically to the isotropic case, yielding the general reversibility relation [Eq.~(\ref{eq:rewinding_iso})]
\begin{align}
\mathcal S_{\mathrm{ord}}
=
\left(
\mathcal S_{\mathrm{rev}}^{*}
\right)^{-1}.
\nonumber
\end{align}

%\subsection*{Structure of the scattering matrix}

Because the temporal interface conditions in chiral media are polarization independent, the scattering matrix of a single chiral slab differs from the isotropic case only through an additional propagation phase originating from optical activity. It can therefore be written as
\begin{align}
\mathcal S
=
e^{i\beta}
\begin{pmatrix}
s & r^{*} \\
r & s^{*}
\end{pmatrix},
\nonumber
\end{align}
where
\begin{align}
\beta
=
\frac{ck\gamma\tau}{n^2-\gamma^2}
\nonumber
\end{align}
is the chirality-induced phase accumulated during a temporal interval of duration $\tau$.
Crucially, $\beta$ enters only as an overall multiplicative factor and leaves the internal $\mathrm{SU}(1,1)$ structure of the scattering matrix unchanged. As a result, the product of two chiral slabs retains the same algebraic form:
\begin{align}
\mathcal S_2 \mathcal S_1
=
e^{i(\beta_{1}+\beta_{2})}
\begin{pmatrix}
s_2 s_1 + r_2^{*} r_1
&
s_2 r_1^{*} + r_2^{*} s_1^{*}
\\
r_2 s_1 + s_2^{*} r_1
&
r_2 r_1^{*} + s_2^{*} s_1^{*}
\end{pmatrix}.
\nonumber
\end{align}
For an ordered temporal sequence, the total scattering matrix can therefore be expressed as
\begin{align}
\mathcal S_{\mathrm{ord}}
=
e^{i\sum_j \beta_{j}}
\begin{pmatrix}
\hat s & \hat r^{*} \\
\hat r & \hat s^{*}
\end{pmatrix},
\nonumber
\end{align}
where the accumulated phase $\sum_j \beta_j$ collects all chirality-induced propagation contributions. Applying the inverse–conjugation symmetry yields the scattering matrix of the reversed sequence,
\begin{align}
\mathcal S_{\mathrm{rev}}
=
\left(\mathcal S_{\mathrm{ord}}^{*}\right)^{-1}
=
e^{i\sum_j \beta_{j}}
\begin{pmatrix}
\hat s & -\hat r \\
-\hat r^{*} & \hat s^{*}
\end{pmatrix}.
\nonumber
\end{align}
It follows that the transmission coefficient remains invariant under temporal reversal,
\begin{align}
\tilde s=s,
\nonumber
\end{align}
whereas the reflection coefficient transforms according to
\begin{align}
\tilde r
=
- r^{*}
e^{2i\sum_j\beta_{j}}.
\nonumber
\end{align}

\section*{Note 5. Time-rewinding conditions in chiral media from temporal reciprocity}

We follow the same steps given in Note 2. Since chiral media obey the same temporal reciprocity identity as isotropic media,
$\mathcal{S}_{\mathrm{rev}}^{*}\,\mathcal{S}_{\mathrm{ord}} = I$, the time-rewinding conditions for a conjugate pair of media $\mathrm B$ and $\mathrm B'$ follow from the requirement
$Q_{\mathrm{B}'} = Q_{\mathrm{B}}^*$, where $Q_{\mathrm B}=\mathcal{S}_{\mathrm{A}\to\mathrm{B}\to\mathrm{A}}$.

Assume first that $\mathrm B$ and $\mathrm B'$ are propagating.
Using the explicit three-stage expression, one obtains
\begin{align}
Q_{\mathrm{B}}
  &= \begin{pmatrix}
       s_{\mathrm{B}\to\mathrm{A}}\,s_{\mathrm{A}\to\mathrm{B}}
       & s_{\mathrm{B}\to\mathrm{A}}\,r_{\mathrm{A}\to\mathrm{B}} \\[4pt]
       r_{\mathrm{B}\to\mathrm{A}}\,s_{\mathrm{A}\to\mathrm{B}}
       & r_{\mathrm{B}\to\mathrm{A}}\,r_{\mathrm{A}\to\mathrm{B}}
     \end{pmatrix} e^{-i\alpha_{\mathrm{B}}}e^{i\beta_{\mathrm{B}}}
   +\begin{pmatrix}
       r_{\mathrm{B}\to\mathrm{A}}\,r_{\mathrm{A}\to\mathrm{B}}
       & r_{\mathrm{B}\to\mathrm{A}}\,s_{\mathrm{A}\to\mathrm{B}} \\[4pt]
       s_{\mathrm{B}\to\mathrm{A}}\,r_{\mathrm{A}\to\mathrm{B}}
       & s_{\mathrm{B}\to\mathrm{A}}\,s_{\mathrm{A}\to\mathrm{B}}
     \end{pmatrix} e^{i\alpha_{\mathrm{B}}}e^{i\beta_{\mathrm{B}}},
\nonumber
\end{align}
where the interface coefficients are real and identical to the isotropic case. The phase factors are
\begin{align}
  \alpha_{\mathrm{B}}=\frac{ckn_{\mathrm B}\tau_{\mathrm B}}{n_{\mathrm B}^2-\gamma_{\mathrm B}^2},\quad
  \beta_{\mathrm{B}}=\frac{ck\gamma_{\mathrm B}\tau_{\mathrm B}}{n_{\mathrm B}^2-\gamma_{\mathrm B}^2},
  \nonumber
\end{align}
with $n_{\mathrm B}=\sqrt{\epsilon_{\mathrm B}}\sqrt{\mu_{\mathrm B}}$.
The condition $Q_{\mathrm{B}'} = Q_{\mathrm{B}}^{*}$ is satisfied if and only if
\begin{align}
  \alpha_{\mathrm{B}'} = -\alpha_{\mathrm{B}},\quad \beta_{\mathrm{B}'} = -\beta_{\mathrm{B}},\quad
  \eta_{\mathrm{B}'} = \eta_{\mathrm{B}}.
  \nonumber
\end{align}
These relations yield the time-rewinding constraints
\begin{align}
  &\frac{\epsilon_{\mathrm{B}'}}{\epsilon_{\mathrm{B}}}
    = \frac{\mu_{\mathrm{B}'}}{\mu_{\mathrm{B}}}=\frac{\gamma_{\mathrm{B}'}}{\gamma_{\mathrm{B}}}, \quad
  \frac{\tau_{\mathrm{B}'}}{\tau_{\mathrm{B}}}
    = \left\lvert\frac{n_{\mathrm{B}^\prime}}{n_{\mathrm{B}}}\right\rvert,\nonumber\\
    & {\mathrm {sgn}}\left(\frac{\epsilon_{\mathrm{B}'}}{\epsilon_{\mathrm{B}}}\right)
    = {\mathrm {sgn}}\left(\frac{\mu_{\mathrm{B}'}}{\mu_{\mathrm{B}}}\right)={\mathrm {sgn}}\left(\frac{\gamma_{\mathrm{B}'}}{\gamma_{\mathrm{B}}}\right)=-1.
\label{eq:trca23}
\end{align}
Thus, conjugate propagating media possess identical impedances but opposite refractive indices and chirality parameters. The slab durations need not coincide; only the ratio specified above is required.

If $\mathrm B$ and $\mathrm B'$ are evanescent, the three-stage matrix becomes
\begin{align}
Q_{\mathrm{B}}
  = \begin{pmatrix}
       s_{\mathrm{B}\to\mathrm{A}}\,s_{\mathrm{A}\to\mathrm{B}}
       & s_{\mathrm{B}\to\mathrm{A}}\,r_{\mathrm{A}\to\mathrm{B}} \\[4pt]
       r_{\mathrm{B}\to\mathrm{A}}\,s_{\mathrm{A}\to\mathrm{B}}
       & r_{\mathrm{B}\to\mathrm{A}}\,r_{\mathrm{A}\to\mathrm{B}}
     \end{pmatrix} e^{\tilde\kappa_{\mathrm{B}}}e^{i\beta_{\mathrm{B}}}
   +\begin{pmatrix}
       r_{\mathrm{B}\to\mathrm{A}}\,r_{\mathrm{A}\to\mathrm{B}}
       & r_{\mathrm{B}\to\mathrm{A}}\,s_{\mathrm{A}\to\mathrm{B}} \\[4pt]
       s_{\mathrm{B}\to\mathrm{A}}\,r_{\mathrm{A}\to\mathrm{B}}
       & s_{\mathrm{B}\to\mathrm{A}}\,s_{\mathrm{A}\to\mathrm{B}}
     \end{pmatrix} e^{-\tilde\kappa_{\mathrm{B}}}e^{i\beta_{\mathrm{B}}},
\nonumber
\end{align}
where $\tilde\kappa_{\mathrm B}$ denotes the attenuation constant and $\beta_{\mathrm B}$ is the chirality-induced phase.
In this regime, the interface coefficients are purely imaginary and satisfy relations such as
$s_{\mathrm{B}\to\mathrm{A}}^{*} = r_{\mathrm{B}\to\mathrm{A}}$.
The condition $Q_{\mathrm{B}'} = Q_{\mathrm{B}}^{*}$ now requires
\begin{align}
  \tilde\kappa_{\mathrm{B}'} = \tilde\kappa_{\mathrm{B}},\quad
  \eta_{\mathrm{B}'} = -\eta_{\mathrm{B}}, \quad \beta_{\mathrm{B}'} = -\beta_{\mathrm{B}}.
  \nonumber
\end{align}
These constraints again reduce to Eq.~\eqref{eq:trca23}.
Thus, conjugate evanescent media possess identical decay rates but opposite impedances and chirality parameters. As in the propagating case, the slab durations are not individually fixed; only the required ratio must be satisfied.

\section*{Note 6. Polarization-exchanging temporal reciprocity in Tellegen media}

At a temporal interface between Tellegen media ($\gamma=0$, $\chi\neq 0$), the scattering matrix has a more intricate structure than in isotropic or chiral media because the effective impedances depend explicitly on polarization. The two circular components are therefore coupled asymmetrically at the interface, and the interfacial scattering must be treated explicitly.

The temporal scattering matrix for an interface from medium $\mathrm A$ to $\mathrm B$ in the circular basis ($D_R$, $D_L$) takes the form
\begin{align}
\mathcal{S}_{\mathrm {A}\to{\mathrm B}}^R=\begin{pmatrix}
  s_{\mathrm {A}\to{\mathrm B}}^R & r_{\mathrm {A}\to{\mathrm B}}^L \\
  r_{\mathrm {A}\to{\mathrm B}}^R & s_{\mathrm {A}\to{\mathrm B}}^L\end{pmatrix},\quad
  \mathcal{S}_{\mathrm {A}\to{\mathrm B}}^L=\begin{pmatrix}
  s_{\mathrm {A}\to{\mathrm B}}^L & r_{\mathrm {A}\to{\mathrm B}}^R \\
  r_{\mathrm {A}\to{\mathrm B}}^L & s_{\mathrm {A}\to{\mathrm B}}^R
\end{pmatrix},
\nonumber
\end{align}
with interface coefficients defined by Eqs.~\eqref{eq:contcond1}–\eqref{eq:contcond2}.
To construct the reversed temporal interface $\mathrm B \to \mathrm A$, we define $\mathcal S_{\mathrm B\to\mathrm A}^R$
and $\mathcal S_{\mathrm B\to\mathrm A}^L$ analogously. A direct substitution of the explicit coefficients shows that
\begin{align}
\rho_{\mathrm B\to\mathrm A}\rho_{\mathrm A\to\mathrm B}=1,
\quad
\nu_{\mathrm B\to\mathrm A}
+
\rho_{\mathrm B\to\mathrm A}\nu_{\mathrm A\to\mathrm B}
=0.
\nonumber
\end{align}
Using these identities, one finds
\begin{align}
\mathcal S_{\mathrm B\to\mathrm A}^R
\mathcal S_{\mathrm A\to\mathrm B}^R=I,
\nonumber
\end{align}
and a similar relation for $L$ channels.
Thus, a Tellegen temporal interface is exactly inverted under reversal of the temporal ordering.
Similarly, two successive Tellegen interfaces concatenate into an effective single interface:
\begin{align}
\mathcal S_{\mathrm B\to\mathrm C}^R
\mathcal S_{\mathrm A\to\mathrm B}^R
=\mathcal S_{\mathrm A\to\mathrm C}^R,
\nonumber
\end{align}
and a similar relation for $L$ channels.
Therefore, despite polarization dependence, Tellegen temporal interfaces retain exact inversion and composition properties.

The symmetry of the interface matrices $\mathcal S_{\mathrm A\to\mathrm B}^R$ and $\mathcal S_{\mathrm A\to\mathrm B}^L$ depends on whether the participating media support propagating or evanescent modes.
If both $\mathrm A$ and $\mathrm B$ are propagating, the explicit interface coefficients yield
\begin{align}
s_{\mathrm A\to\mathrm B}^R=
\left(s_{\mathrm A\to\mathrm B}^L\right)^*,
\quad
r_{\mathrm A\to\mathrm B}^R
=
\left(r_{\mathrm A\to\mathrm B}^L\right)^*.
\label{eq:tellwa}
\end{align}
If $\mathrm B$ is evanescent while $\mathrm A$ is propagating,
\begin{align}
s_{\mathrm A\to\mathrm B}^R=
\left(r_{\mathrm A\to\mathrm B}^L\right)^*,
\quad
r_{\mathrm A\to\mathrm B}^R
=
\left(s_{\mathrm A\to\mathrm B}^L\right)^*.
\label{eq:tellwb}
\end{align}
If $\mathrm A$ is evanescent and $\mathrm B$ propagating,
\begin{align}
s_{\mathrm{A}\to\mathrm{B}}^{R}=\left(r_{\mathrm{A}\to\mathrm{B}}^{R}\right)^{*},
\quad
s_{\mathrm{A}\to\mathrm{B}}^{L}=\left(r_{\mathrm{A}\to\mathrm{B}}^{L}\right)^{*}.
\label{eq:tellwc}
\end{align}
These cases can be compactly summarized using the permutation matrix $M$:
\begin{align}
\left(\mathcal S_{\mathrm A\to\mathrm B}^R\right)^*
=
\begin{cases}
M \mathcal S_{\mathrm A\to\mathrm B}^R M,
& \text{propagating A, B}, \\
\mathcal S_{\mathrm A\to\mathrm B}^R M,
& \text{propagating A, evanescent B}, \\
M \mathcal S_{\mathrm A\to\mathrm B}^R,
& \text{evanescent A, propagating B}.
\end{cases}
\label{eq:tellw}
\end{align}

Consider now a three-stage process $\mathrm A\to\mathrm B\to\mathrm C$, with $\mathrm A$ and $\mathrm C$ propagating.
Unlike isotropic media, Tellegen media do not satisfy a simple inverse–conjugation relation within a single circular channel. However, reciprocity is restored in the circular basis.
For both propagating and evanescent intermediate media $\mathrm B$, Eq.~(\ref{eq:tellw}) allows the three-stage matrix to be written in the unified form
\begin{align}
\mathcal S_{\mathrm A\to\mathrm B\to\mathrm C}^R
=
\left[
\left(
M
\mathcal S_{\mathrm B\to\mathrm A}^R
M
\mathcal P_{\mathrm B}
M
\mathcal S_{\mathrm C\to\mathrm B}^R
M
\right)^*
\right]^{-1}.
\nonumber
\end{align}
Importantly,
\begin{align}
M \mathcal S_{\mathrm B\to\mathrm A}^R M
&=
\begin{pmatrix}
s_{\mathrm B\to\mathrm A}^L & r_{\mathrm B\to\mathrm A}^R \\
r_{\mathrm B\to\mathrm A}^L & s_{\mathrm B\to\mathrm A}^R
\end{pmatrix}\nonumber\\
&=\mathcal S_{\mathrm B\to\mathrm A}^L,
\nonumber
\end{align}
which has the same algebraic structure as the interface matrix of the opposite circular channel.
Thus, temporal reciprocity in Tellegen media intrinsically couples opposite circular polarizations: the RCP channel of the ordered sequence is mapped to the LCP channel of the reversed sequence, and vice versa. The extension from a three-stage process to an arbitrary multistage temporal modulation proceeds identically to the isotropic case. The resulting reciprocity relation therefore takes the form
\begin{align}
\mathcal S_{\mathrm{ord}}^{R}
=
[
(
\mathcal S_{\mathrm{rev}}^{L}
)^*
]^{-1},
\label{eq:trev_refined}
\end{align}
where superscripts denote circular channels.
Equivalently, ordered-sequence scattering in the $D_R$ channel equals reversed-sequence scattering in the same channel with the Tellegen parameter $\chi$ sign-reversed.

Within each circular channel, the scattering matrix preserves the $\mathrm{SU}(1,1)$ structure,
\begin{align}
\mathcal S_{\mathrm{ord}}^R=
\begin{pmatrix}
s^R & \left(r^R\right)^* \\
r^R & \left(s^R\right)^*
\end{pmatrix}, \quad
\mathcal S_{\mathrm{ord}}^L=
\begin{pmatrix}
s^L & \left(r^L\right)^* \\
r^L & \left(s^L\right)^*
\end{pmatrix}.
\label{eq:tellw2}
\end{align}
Combining Eqs.~(\ref{eq:trev_refined}) and (\ref{eq:tellw2}), the reversed-sequence matrix in the $L$ channel becomes
\begin{align}
   \mathcal S_{\mathrm{rev}}^{L}= \begin{pmatrix}
s^R & -r^R \\
-\left(r^R\right)^* & \left(s^R\right)^*
\end{pmatrix}.
\nonumber
\end{align}
Comparison with $\mathcal S_{\mathrm{ord}}^R$ shows that the transmission amplitudes coincide,
\begin{align}
s^R = \tilde s^L,
\nonumber
\end{align}
whereas the reflection amplitudes differ by a $\pi$ phase shift (up to complex conjugation),
\begin{align}
r^R = -(\tilde r^L)^*.
\nonumber
\end{align}
Thus, as in isotropic and chiral media, transmission is preserved under temporal reversal, while reflection acquires a sign change; the distinctive feature of Tellegen media is the exchange of circular channels.

\section*{Note 7. Time-rewinding conditions in Tellegen media from polarization-exchanging temporal reciprocity}

Time-rewinding conditions in Tellegen media are derived by the same block-factorization strategy used for isotropic and chiral media, with one essential modification: the symmetry relating ordered and reversed sequences is the \emph{channel-exchanging} temporal reciprocity,
\begin{align}
\mathcal S_{\mathrm{ord}}^{R}
=
[(\mathcal S_{\mathrm{rev}}^{L})^{*}]^{-1},
\nonumber
\end{align}
rather than the channel-preserving inverse--conjugation relation.

%\subsection*{Block factorization and rewinding criterion}

Consider the ordered sequence
$\mathrm A\to \mathrm B\to \mathrm C\to \mathrm A$
evaluated in the RCP sector. Its scattering matrix factorizes as
\begin{align}
\mathcal S_{\mathrm{ord}}^{R}=Q_{\mathrm C}^{R}\,Q_{\mathrm B}^{R},
\nonumber
\end{align}
where
\begin{align}
Q_{\mathrm B}^{R}
=
\mathcal S_{\mathrm{B}\to\mathrm A}^{R}\,\mathcal P_{\mathrm B}\,\mathcal S_{\mathrm A\to\mathrm B}^{R},
\quad
Q_{\mathrm C}^{R}
=
\mathcal S_{\mathrm{C}\to\mathrm A}^{R}\,\mathcal P_{\mathrm C}\,\mathcal S_{\mathrm A\to\mathrm C}^{R}.
\nonumber
\end{align}
For the reversed ordering $\mathrm A\to \mathrm C\to \mathrm B\to \mathrm A$ evaluated in the LCP sector, one similarly obtains
\begin{align}
\mathcal S_{\mathrm{rev}}^{L}=Q_{\mathrm B}^{L}\,Q_{\mathrm C}^{L}.
\nonumber
\end{align}
Substitution into the temporal reciprocity relation yields
\begin{align}
\left(Q_{\mathrm B}^{L}\right)^{*}\,\left(Q_{\mathrm C}^{L}\right)^{*}\,Q_{\mathrm C}^{R}\,Q_{\mathrm B}^{R}=I.
\nonumber
\end{align}

We now construct a rewinding sequence of the form
\[
\mathrm A\to \mathrm B\to \mathrm C\to \mathrm C'\to \mathrm B'\to \mathrm A,
\]
where $(\mathrm B,\mathrm B')$ and $(\mathrm C,\mathrm C')$ form generalized conjugate pairs. The total evolution operator reads
\[
\mathcal S = Q_{\mathrm B'}^{R}\,Q_{\mathrm C'}^{R}\,Q_{\mathrm C}^{R}\,Q_{\mathrm B}^{R}.
\]
Perfect temporal rewinding ($\mathcal S = I$) is achieved provided
\begin{align}
Q_{\mathrm B'}^{R}=\left(Q_{\mathrm B}^{L}\right)^{*},
\quad
Q_{\mathrm C'}^{R}=\left(Q_{\mathrm C}^{L}\right)^{*}.
\label{eq:rewind_conditiontw}
\end{align}
In the following, we derive the explicit constraints on the conjugate Tellegen parameters implied by Eq.~\eqref{eq:rewind_conditiontw}.

\subsection*{Propagating conjugate media}

Assume that both $\mathrm B$ and $\mathrm B'$ are propagating.
The three-stage block in the RCP sector,
$Q_{\mathrm B'}^{R}=\mathcal S_{\mathrm A\to\mathrm B'\to\mathrm A}^{R}$, can be written as
\begin{align}
Q_{\mathrm{B}^\prime}^{R}
&=
\begin{pmatrix}
s_{\mathrm{B}^\prime\to\mathrm{A}}^{R}s_{\mathrm{A}\to\mathrm{B}^\prime}^{R}
&
s_{\mathrm{B}^\prime\to\mathrm{A}}^{R}r_{\mathrm{A}\to\mathrm{B}^\prime}^{L}
\\[4pt]
r_{\mathrm{B}^\prime\to\mathrm{A}}^{R}s_{\mathrm{A}\to\mathrm{B}^\prime}^{R}
&
r_{\mathrm{B}^\prime\to\mathrm{A}}^{R}r_{\mathrm{A}\to\mathrm{B}^\prime}^{L}
\end{pmatrix} e^{-i\hat\phi_{\mathrm{B}^\prime}}
+
\begin{pmatrix}
r_{\mathrm{B}^\prime\to\mathrm{A}}^{L}r_{\mathrm{A}\to\mathrm{B}^\prime}^{R}
&
r_{\mathrm{B}^\prime\to\mathrm{A}}^{L}s_{\mathrm{A}\to\mathrm{B}^\prime}^{L}
\\[4pt]
s_{\mathrm{B}^\prime\to\mathrm{A}}^{L}r_{\mathrm{A}\to\mathrm{B}^\prime}^{R}
&
s_{\mathrm{B}^\prime\to\mathrm{A}}^{L}s_{\mathrm{A}\to\mathrm{B}^\prime}^{L}
\end{pmatrix} e^{i\hat\phi_{\mathrm{B}^\prime}},
\label{eq:xxx1}
\end{align}
where
\begin{align}
\hat\phi_{\mathrm B^\prime}=\frac{ck\tau_{\mathrm B^\prime}}{n_{0\mathrm B^\prime}}.
\nonumber
\end{align}
The corresponding LCP block for $\mathrm B$ reads
\begin{align}
Q_{\mathrm{B}}^{L}
&=
\begin{pmatrix}
s_{\mathrm{B}\to\mathrm{A}}^{L}s_{\mathrm{A}\to\mathrm{B}}^{L}
&
s_{\mathrm{B}\to\mathrm{A}}^{L}r_{\mathrm{A}\to\mathrm{B}}^{R}
\\[4pt]
r_{\mathrm{B}\to\mathrm{A}}^{L}s_{\mathrm{A}\to\mathrm{B}}^{L}
&
r_{\mathrm{B}\to\mathrm{A}}^{L}r_{\mathrm{A}\to\mathrm{B}}^{R}
\end{pmatrix} e^{-i\hat\phi_{\mathrm{B}}}
+
\begin{pmatrix}
r_{\mathrm{B}\to\mathrm{A}}^{R}r_{\mathrm{A}\to\mathrm{B}}^{L}
&
r_{\mathrm{B}\to\mathrm{A}}^{R}s_{\mathrm{A}\to\mathrm{B}}^{R}
\\[4pt]
s_{\mathrm{B}\to\mathrm{A}}^{R}r_{\mathrm{A}\to\mathrm{B}}^{L}
&
s_{\mathrm{B}\to\mathrm{A}}^{R}s_{\mathrm{A}\to\mathrm{B}}^{R}
\end{pmatrix} e^{i\hat\phi_{\mathrm{B}}}.
\nonumber
\end{align}
For propagating $\mathrm A$ and $\mathrm B$, the coefficient relations of Eq.~\eqref{eq:tellwa} give
\begin{align}
\left(Q_{\mathrm B}^{L}\right)^{*}
&=
\begin{pmatrix}
s_{\mathrm{B}\to\mathrm{A}}^{R}s_{\mathrm{A}\to\mathrm{B}}^{R}
&
s_{\mathrm{B}\to\mathrm{A}}^{R}r_{\mathrm{A}\to\mathrm{B}}^{L}
\\[4pt]
r_{\mathrm{B}\to\mathrm{A}}^{R}s_{\mathrm{A}\to\mathrm{B}}^{R}
&
r_{\mathrm{B}\to\mathrm{A}}^{R}r_{\mathrm{A}\to\mathrm{B}}^{L}
\end{pmatrix} e^{i\hat\phi_{\mathrm{B}}}
+
\begin{pmatrix}
r_{\mathrm{B}\to\mathrm{A}}^{L}r_{\mathrm{A}\to\mathrm{B}}^{R}
&
r_{\mathrm{B}\to\mathrm{A}}^{L}s_{\mathrm{A}\to\mathrm{B}}^{L}
\\[4pt]
s_{\mathrm{B}\to\mathrm{A}}^{L}r_{\mathrm{A}\to\mathrm{B}}^{R}
&
s_{\mathrm{B}\to\mathrm{A}}^{L}s_{\mathrm{A}\to\mathrm{B}}^{L}
\end{pmatrix} e^{-i\hat\phi_{\mathrm{B}}}.
\label{eq:xxx2}
\end{align}
Comparison of Eqs.~\eqref{eq:xxx1} and \eqref{eq:xxx2} shows that $Q_{\mathrm B'}^{R}=\left(Q_{\mathrm B}^{L}\right)^{*}$ is equivalent to
\begin{align}
    \hat\phi_{\mathrm B'}=-\hat\phi_{\mathrm B},
\quad s_{\mathrm{A}\to\mathrm{B}}^{R/L}=s_{\mathrm{A}\to\mathrm{B^\prime}}^{R/L},\quad
r_{\mathrm{A}\to\mathrm{B}}^{R/L}=r_{\mathrm{A}\to\mathrm{B^\prime}}^{R/L},
\nonumber
\end{align}
which gives
\begin{align}
\hat\phi_{\mathrm B'}=-\hat\phi_{\mathrm B},
\quad
\rho_{\mathrm A\to\mathrm B'}=\rho_{\mathrm A\to\mathrm B},
\quad
\nu_{\mathrm A\to\mathrm B'}=\nu_{\mathrm A\to\mathrm B}.
\nonumber
\end{align}
These relations yield the rewinding constraints
\begin{align}
\frac{\epsilon_{\mathrm{B}'}}{\epsilon_{\mathrm{B}}}
=
\frac{\mu_{\mathrm{B}'}}{\mu_{\mathrm{B}}}
=
\frac{\chi_{\mathrm{B}'}}{\chi_{\mathrm{B}}},
\quad
\frac{\tau_{\mathrm{B}'}}{\tau_{\mathrm{B}}}
=
\left|\frac{n_{0\mathrm{B}'}}{n_{0\mathrm{B}}}\right|,
\label{eq:trca11a}
\end{align}
together with
\begin{align}
\operatorname{sgn}\!\left(\frac{\epsilon_{\mathrm{B}'}}{\epsilon_{\mathrm{B}}}\right)
=
\operatorname{sgn}\!\left(\frac{\mu_{\mathrm{B}'}}{\mu_{\mathrm{B}}}\right)
=
\operatorname{sgn}\!\left(\frac{\chi_{\mathrm{B}'}}{\chi_{\mathrm{B}}}\right)
=-1,
\label{eq:trca11b}
\end{align}
where $n_{0\mathrm B}=\sqrt{\epsilon_{\mathrm B}\mu_{\mathrm B}-\chi_{\mathrm B}^{2}}$.

\subsection*{Evanescent conjugate media}

If $\mathrm B$ and $\mathrm B'$ are evanescent, the three-stage block becomes
\begin{align}
Q_{\mathrm{B^\prime}}^{R}
&=
\begin{pmatrix}
s_{\mathrm{B^\prime}\to\mathrm{A}}^{R}s_{\mathrm{A}\to\mathrm{B^\prime}}^{R}
&
s_{\mathrm{B^\prime}\to\mathrm{A}}^{R}r_{\mathrm{A}\to\mathrm{B^\prime}}^{L}
\\[4pt]
r_{\mathrm{B^\prime}\to\mathrm{A}}^{R}s_{\mathrm{A}\to\mathrm{B^\prime}}^{R}
&
r_{\mathrm{B^\prime}\to\mathrm{A}}^{R}r_{\mathrm{A}\to\mathrm{B^\prime}}^{L}
\end{pmatrix} e^{\hat\kappa_{\mathrm{B^\prime}}}
+
\begin{pmatrix}
r_{\mathrm{B^\prime}\to\mathrm{A}}^{L}r_{\mathrm{A}\to\mathrm{B^\prime}}^{R}
&
r_{\mathrm{B^\prime}\to\mathrm{A}}^{L}s_{\mathrm{A}\to\mathrm{B^\prime}}^{L}
\\[4pt]
s_{\mathrm{B^\prime}\to\mathrm{A}}^{L}r_{\mathrm{A}\to\mathrm{B^\prime}}^{R}
&
s_{\mathrm{B^\prime}\to\mathrm{A}}^{L}s_{\mathrm{A}\to\mathrm{B^\prime}}^{L}
\end{pmatrix} e^{-\hat\kappa_{\mathrm{B^\prime}}},
\nonumber
\end{align}
with
\begin{align}
\hat\kappa_{\mathrm B^\prime}=\frac{ck\tau_{\mathrm B^\prime}}{|n_{0\mathrm B^\prime}|}.
\nonumber
\end{align}
Using Eqs.~\eqref{eq:tellwb}--\eqref{eq:tellwc}, one finds
\begin{align}
\left(Q_{\mathrm B}^{L}\right)^{*}
&=
\begin{pmatrix}
r_{\mathrm{B}\to\mathrm{A}}^{L}r_{\mathrm{A}\to\mathrm{B}}^{R}
&
r_{\mathrm{B}\to\mathrm{A}}^{L}s_{\mathrm{A}\to\mathrm{B}}^{L}
\\[4pt]
s_{\mathrm{B}\to\mathrm{A}}^{L}r_{\mathrm{A}\to\mathrm{B}}^{R}
&
s_{\mathrm{B}\to\mathrm{A}}^{L}s_{\mathrm{A}\to\mathrm{B}}^{L}
\end{pmatrix} e^{\hat\kappa_{\mathrm{B}}}
+
\begin{pmatrix}
s_{\mathrm{B}\to\mathrm{A}}^{R}s_{\mathrm{A}\to\mathrm{B}}^{R}
&
s_{\mathrm{B}\to\mathrm{A}}^{R}r_{\mathrm{A}\to\mathrm{B}}^{L}
\\[4pt]
r_{\mathrm{B}\to\mathrm{A}}^{R}s_{\mathrm{A}\to\mathrm{B}}^{R}
&
r_{\mathrm{B}\to\mathrm{A}}^{R}r_{\mathrm{A}\to\mathrm{B}}^{L}
\end{pmatrix} e^{-\hat\kappa_{\mathrm{B}}}.
\nonumber
\end{align}
Matching this with $Q_{\mathrm B'}^{R}$ yields
\begin{align}
&\hat\kappa_{\mathrm B'}=\hat\kappa_{\mathrm B},
\quad
\rho_{\mathrm A\to\mathrm B'}=-\rho_{\mathrm A\to\mathrm B},
\quad
\nu_{\mathrm A\to\mathrm B'}=-\nu_{\mathrm A\to\mathrm B},\nonumber\\
&\rho_{\mathrm B'\to\mathrm A}=-\rho_{\mathrm B\to\mathrm A},
\quad
\nu_{\mathrm B'\to\mathrm A}=\nu_{\mathrm B\to\mathrm A}.
\nonumber
\end{align}
These again reduce to the parameter constraints
\eqref{eq:trca11a}--\eqref{eq:trca11b}, now with $n_{0\mathrm B}$ and $n_{0\mathrm B'}$ purely imaginary. The conjugate partner thus possesses identical decay rate but opposite impedance.

\section*{Note 8. Numerical results confirming temporal reciprocity}

We present representative numerical examples that confirm the reciprocity relations for isotropic and chiral media derived in the main text and Supplemental Information; the Tellegen case is discussed in the main text.
All calculations were performed using the invariant imbedding
method developed in Ref.~[12] of the main text and extended
here to bi-isotropic media.

\subsection*{Isotropic media}

For isotropic media, temporal sequence reversal yields
\begin{align}
\tilde s = s,\quad \tilde r = -\,r^{*},
\label{eq:recipro}
\end{align}
so that transmission amplitudes are invariant under reversal, while reflection amplitudes retain the same imaginary part and acquire a sign reversal of the real part.

As a representative example, we consider a piecewise-constant temporal modulation composed of three segments of equal duration $\tau_0$ (total duration $3\tau_0$). The permittivity and permeability are
\begin{align}
\epsilon(t)&=
\begin{cases}
2, & 0\le t/\tau_0\le1,\\
3, & 1< t/\tau_0\le2,\\
-1, & 2< t/\tau_0\le3,\\
1, & \text{otherwise},
\end{cases}
&
\mu(t)&=
\begin{cases}
1.5, & 0\le t/\tau_0\le1,\\
-0.5, & 1< t/\tau_0\le2,\\
-2, & 2< t/\tau_0\le3,\\
1, & \text{otherwise}.
\end{cases}
\nonumber
\end{align}
For the reversed configuration, the parameters are arranged in opposite temporal order. To demonstrate that temporal reciprocity is an intrinsic algebraic symmetry of the scattering matrix and does not require one-to-one matching of segment durations or material values, we construct a reversed sequence with modified parameters and a different subdivision (while keeping the total duration $3\tau_0$, but using nonuniform sub-intervals):
\begin{align}
\epsilon(t)&=
\begin{cases}
-1.5, & 0\le t/\tau_0<1.5,\\
1.5, & 1.5\le t/\tau_0<2,\\
2, & 2\le t/\tau_0\le3,\\
1, & \text{otherwise},
\end{cases}
&
\mu(t)&=
\begin{cases}
-3, & 0\le t/\tau_0<1.5,\\
-0.25, & 1.5\le t/\tau_0<2,\\
1.5, & 2\le t/\tau_0\le3,\\
1, & \text{otherwise}.
\end{cases}
\nonumber
\end{align}

\begin{figure}
  \centering
  \includegraphics[width=12cm]{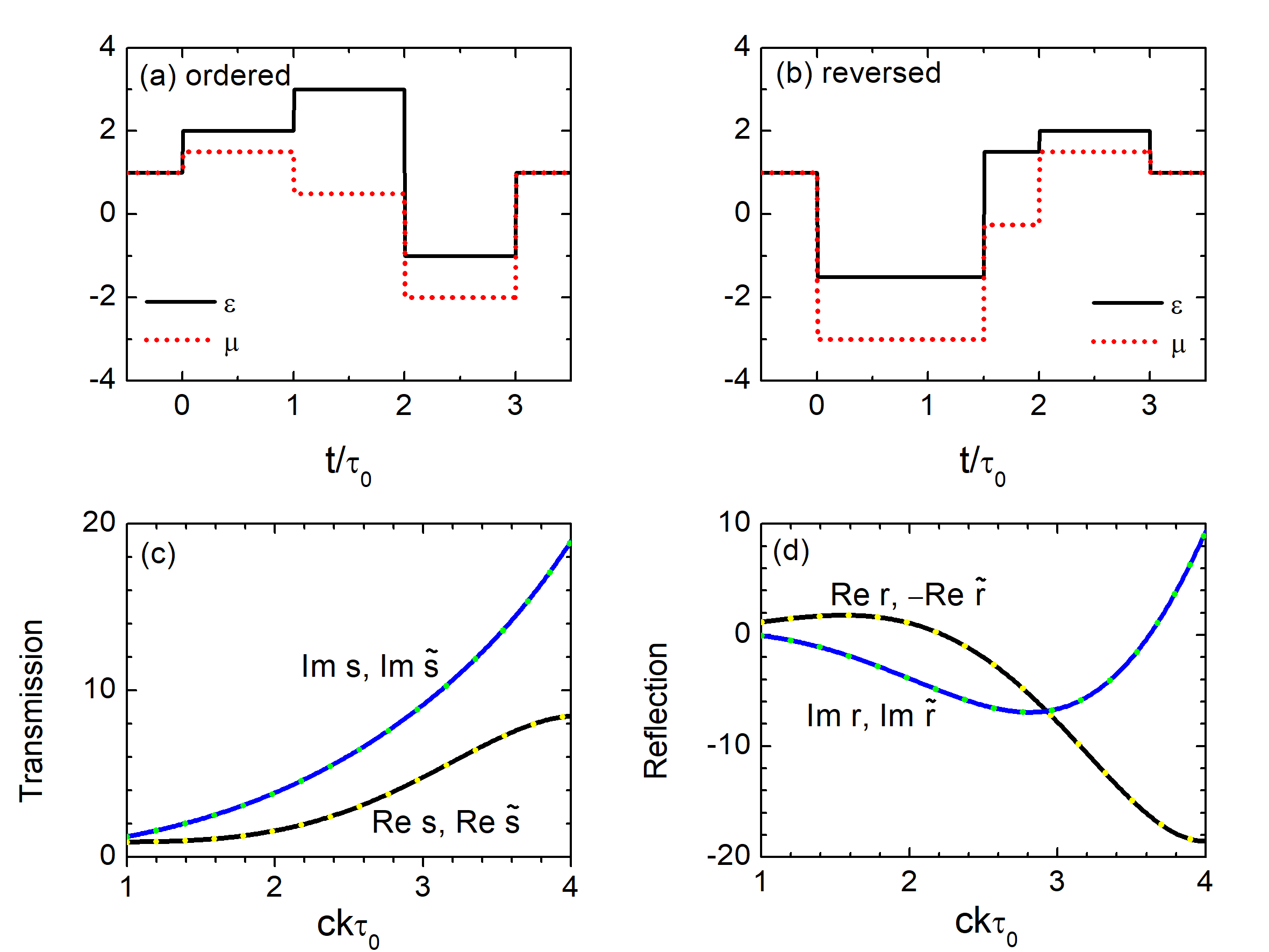}
  \caption{Temporal reciprocity in isotropic media.
(a),(b) Temporal profiles of $\epsilon(t)$ and $\mu(t)$ for the ordered and reversed sequences.
(c),(d) Scattering amplitudes versus normalized wavenumber $ck\tau_0$. Transmission is invariant under reversal ($\tilde s=s$), while reflection obeys $\tilde r=-r^{*}$, yielding identical imaginary parts and real parts of equal magnitude with opposite sign.}
  \label{fig:num_iso}
\end{figure}

Figure~\ref{fig:num_iso} displays the temporal profiles and the corresponding scattering amplitudes as functions of the normalized wavenumber $ck\tau_0$. The numerical results verify Eq.~\eqref{eq:recipro} across the entire plotted range: $\tilde s$ coincides with $s$, while $\tilde r=-r^{*}$.

\subsection*{Chiral media}

We next verify temporal reciprocity for chiral media using a continuously varying modulation.
For the ordered sequence of duration $\tau_0$, we choose
\begin{align}
&\epsilon(t)=\begin{cases}
1+2\frac{t}{\tau_0}, & 0\le t/\tau_0\le1,\\
1, & \text{otherwise},
\end{cases}
\qquad
\mu(t)=\begin{cases}
1-0.5\frac{t}{\tau_0}, & 0\le t/\tau_0\le1,\\
1, & \text{otherwise},
\end{cases}\nonumber\\
&\gamma(t)=\begin{cases}
0.5, & 0\le t/\tau_0\le1,\\
0, & \text{otherwise}.
\end{cases}
\nonumber
\end{align}
For the reversed sequence, the parameters are applied in reverse temporal order, with total duration $2\tau_0$:
\begin{align}
&\epsilon(t)=\begin{cases}
6-2\frac{t}{\tau_0}, & 0\le t/\tau_0\le2,\\
1, & \text{otherwise},
\end{cases}
\qquad
\mu(t)=\begin{cases}
1+0.5\frac{t}{\tau_0}, & 0\le t/\tau_0\le2,\\
1, & \text{otherwise},
\end{cases}
\nonumber\\
&\gamma(t)=\begin{cases}
1, & 0\le t/\tau_0\le2,\\
0, & \text{otherwise}.
\end{cases}
\nonumber
\end{align}

As shown in Note 4, chirality contributes an additional \emph{global} phase factor $e^{i\sum_j \beta_j}$ to the scattering matrix. Consequently, transmission remains invariant under temporal reversal, whereas reflection acquires an additional chirality-induced phase,
\begin{align}
\tilde s = s,
\qquad
\tilde r = -\,e^{2i\Phi}\, r^{*},
\label{eq:num_chiral_rel}
\end{align}
where $\Phi$ ($=\sum_j \beta_j$) denotes the net chirality phase accumulated across the modulation.
For the present profile,
\begin{align}
\Phi=ck\tau_0\,\int_0^1\frac{0.5}{(1+2x)(1-0.5x)-0.5^2}\,dx\approx 0.439 \,ck\tau_0.
\nonumber
\end{align}

\begin{figure}
  \centering
  \includegraphics[width=12cm]{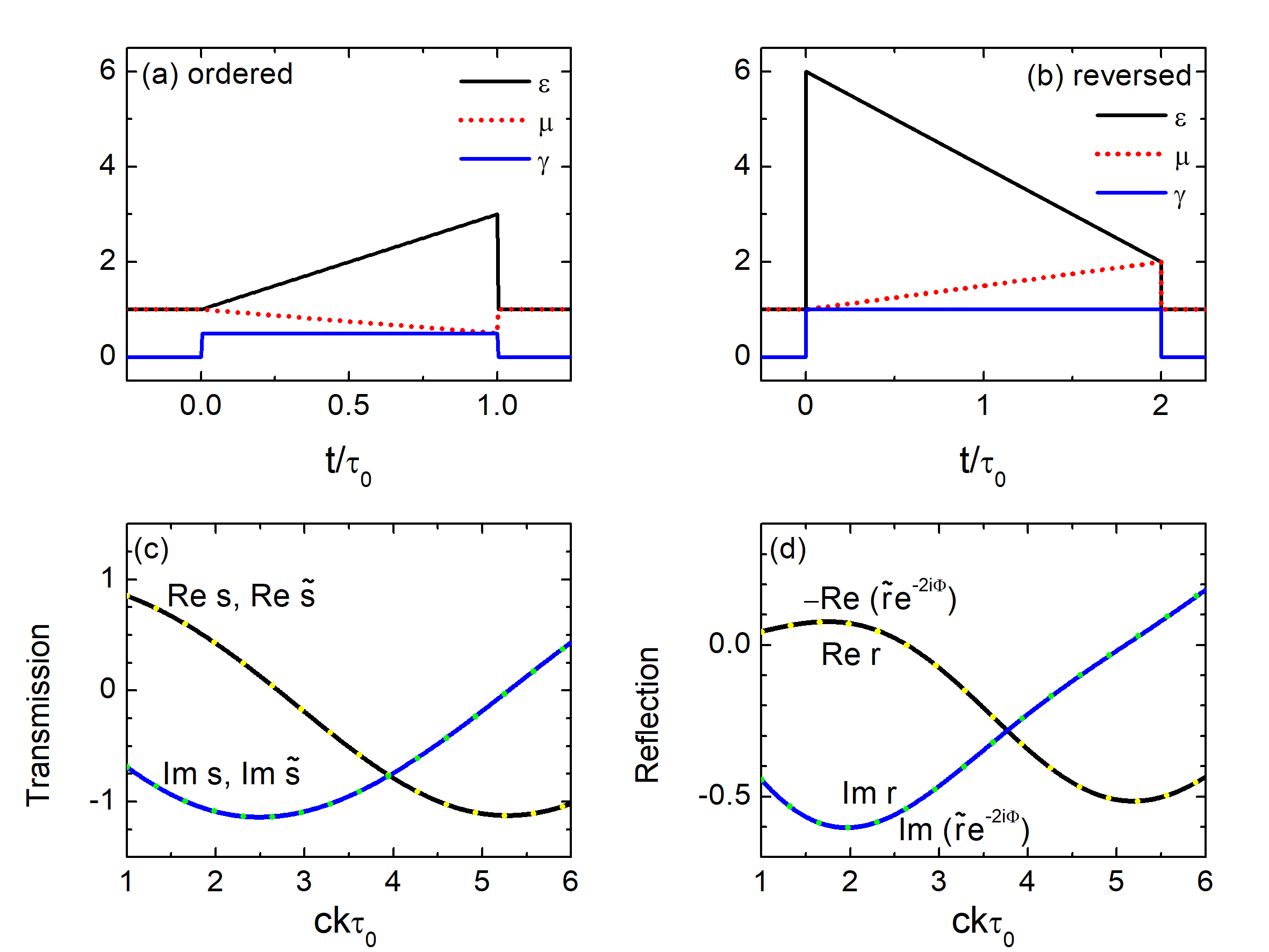}
  \caption{Temporal reciprocity in chiral media.
(a),(b) Temporal profiles of $\epsilon(t)$, $\mu(t)$, and $\gamma(t)$ for the ordered and reversed sequences.
(c),(d) Scattering amplitudes versus $ck\tau_0$. Transmission is invariant under reversal. Reflection obeys $\tilde r=-e^{2i\Phi}r^{*}$, with $\Phi\approx 0.439\,ck\tau_0$.}
  \label{fig:num_chiral}
\end{figure}

Figure~\ref{fig:num_chiral} confirms Eq.~\eqref{eq:num_chiral_rel}: the transmission amplitudes coincide for the ordered and reversed sequences, while the reflection amplitudes satisfy $\tilde r=-e^{2i\Phi} r^{*}$ with $\Phi\approx 0.439\, ck\tau_0$.

\end{document}